\RequirePackage{fix-cm}
\documentclass[twocolumn,epjc3]{svjour3}
\smartqed  % flush right qed marks, e.g. at end of proof
\RequirePackage{graphicx}

\usepackage{amssymb,graphics}
\usepackage{epsfig,subfigure}
\usepackage{color}

\journalname{Eur. Phys. J. C}
\begin{document}

\title{Warped Brane Worlds in Critical Gravity}
%\subtitle{Do you have a subtitle?\\ If so, write it here}

%\titlerunning{Short form of title}        % if too long for running head

%\author{First Author\thanksref{e1,addr1}
%        \and
%        Second Author\thanksref{e2,addr2,addr3} %etc.
%}

\author{Yi Zhong\thanksref{e1,addr1}
        \and Feng-Wei Chen\thanksref{e2,addr1}
         \and Qun-Ying Xie\thanksref{e3,addr1,addr2}
         \and Yu-Xiao Liu\thanksref{e4,addr1}
        }

%\thankstext{t1}{Grants or other notes
%about the article that should go on the front page should be
%placed here. General acknowledgments should be placed at the end of the article.
\thankstext{e1}{e-mail:zhongy13@lzu.edu.cn}
\thankstext{e2}{e-mail:chenfw10@lzu.edu.cn}
\thankstext{e3}{e-mail:xieqy@lzu.edu.cn}
\thankstext{e4}{e-mail:liuyx@lzu.edu.cn}

%\authorrunning{Short form of author list} % if too long for running head

\institute{Institute of Theoretical Physics,
            Lanzhou University, Lanzhou 730000,
            People's Republic of China \label{addr1}
           \and
           School of Information Science and Engineering,
            Lanzhou University, Lanzhou 730000,
            People's Republic of China \label{addr2}
}

\date{Received: date / Accepted: date}
% The correct dates will be entered by the editor

\maketitle

\begin{abstract}
We investigate the brane models in arbitrary dimensional critical gravity presented in
[Phys. Rev. Lett. 106, 181302 (2011)]. For the model of the thin branes with codimension one, the Gibbons-Hawking surface term and the junction conditions are derived, with which the analytical solutions for the flat, AdS, and dS branes are obtained at
the critical point of the critical gravity. It is found that all these branes are embedded in an AdS$_{n}$ spacetime, but, in general, the effective cosmological constant $\Lambda$ of the AdS$_{n}$ spacetime is not equal to the naked one $\Lambda_0$ in the critical gravity, which can be positive, zero, and negative. Another interesting result is that the brane tension can also be positive, zero, or negative, depending on the symmetry of the thin brane and the values of the parameters of the theory, which is very different from the case in general relativity.
It is shown that the mass hierarchy problem can be solved in the braneworld model in the higher-derivative critical gravity.
We also study the thick brane model and find analytical and numerical solutions of the flat, AdS, and dS branes. It is find that some branes will have inner structure when some parameters of the theory are larger than their critical values, which may result in resonant KK modes for some bulk matter fields.
The flat branes with positive energy density and AdS branes with negative energy density are embedded in an $n$-dimensional AdS spacetime, while the dS branes with positive energy density are embedded in an $n$-dimensional Minkowski one.
% \PACS{PACS code1 \and PACS code2 \and more}
% \subclass{MSC code1 \and MSC code2 \and more}
\end{abstract}

\section{Introduction}
The idea that the spacetime has more than four dimensions and our universe
is a brane or domain wall embedded in higher dimensional spacetime \cite{Rubakov:1983bb,Rubakov:1983bz,Visser:1985qm,Squires:1985aq,Randjbar-Daemi:1986,Antoniadis:1990ew} has been proposed for a long time and discussed extendedly.
It is believed that the braneworld scenario can supply new insights for solving
the gauge hierarchy problem and the cosmological constant problem
\cite{Randall:1999ee,Randall:1999vf,Lykken:1999nb,ArkaniHamed:1998rs,Antoniadis:1998ig,Gogberashvili:1998vx,ArkaniHamed:2000eg,Kachru:2000hf,Kehagias:2004fb,Bouhmadi-Lopez:2013gqa,Dutra:2013jea}.
%These branes are all in the frame of general relativity.

There are many discussions about branes both in the frames of general gravity and modified gravities. In Refs. \cite{Yang:2012dd,Bogdanos:2006qw,Liu:2012gv,Guo:2011wr,Ahmed:2012nh,Kar:2013tsa}, non-minimal coupling branes in scalar-tensor gravity were discussed and the mass hierarchy problem can be solved in scalar-tensor thin branes model \cite{Yang:2012dd}.
A brane model in the Recently presented EiBI gravity theory were
constructed in Ref. \cite{Liu:2012rc} and it was found that the four-dimensional Einstein gravity can be recovered on the brane at low energy. Branes in spacetime with torsion were investigated in
Ref \cite{Yang:2012hu,Maier:2012ga} and it was shown that in the $f(T)$ gravity that the torsion of spacetime can effect the inner structure of branes \cite{Yang:2012hu}. Reference \cite{Nozari:2012qi} investigated braneworld teleparallel gravity.
For brane models in higher-derivative gravity there are also many references, see for examples Refs. \cite{Giovannini:2001ta,Nojiri:2001ae,Cho:2001nf,Afonso:2007zz,Dzhunushaliev:2009dt,Zhong:2010ae,Bazeia:2013uva,German:2013sk,Bazeia:2013oha,Neupane:2001st,Cho:2001su}.

%, one motivation of which comes from quantum gravity

In this paper, we are interested in brane solutions in the framework of higher-derivative gravities. As is known, general relativity is a non-renormalizable
theory and it is suffered the singularity problem as well as other problems.
In a quantum gravity theory or the low energy effective theory of string theory, higher-order curvature terms would be added to the Einstein-Hilbert action.
The principal candidates for such corrections are contracted quadratic products of the Riemann
curvature tensor. With this kind of corrections, the most general correction terms has the form of
$\alpha' R^{2}+\beta' R_{MN}R^{MN}+\gamma' R_{MNPQ}R^{MNPQ}$, or $\alpha R^{2}+\beta R_{MN}R^{MN}+\gamma \mathcal{L}_{\mathrm{GB}}$, where
\begin{eqnarray}
   \mathcal{L}_{\mathrm{GB}}=R_{MNPQ}R^{MNPQ}-4R_{MN}R^{MN}+R^{2}
    \label{GBterm}
\end{eqnarray}
is the Gauss-Bonnet term and it is topological invariant in four dimensions.
{
Gravity theories with this kind of corrections were studied in detail in
Ref. \cite{Stelle:1977ry}.
}

Nevertheless, an action with quadratic curvature terms implies that the field equations contain the fourth derivations of the metric and thus would lead to massive ghost-like graviton. Recently, it was shown in Ref. \cite{Lu:2011zk} that the massive scalar mode can be
eliminated and the massive ghost-like graviton becomes massless when the parameters of the quadratic
curvature terms satisfy the critical condition. The corresponding theory is called critical gravity. It was generalized to higher
dimensions in Ref. \cite{Deser:2011xc}. See Ref. \cite{Kan:2012mn,Kan:2013moa,Grumiller:2013at,Johansson:2012fs,Porrati:2011ku} for works related to critical gravity.

In general, it is very hard to solve analytically the Einstein field equations of a higher-derivative  gravity for a system with matter fields. However, for a { co-dimension-1 brane system with the metric $ds^2 = \mathrm{e}^{2A(y)} \hat{g}_{\mu\nu}$ $ dx^{\mu} dx^{\nu} + dy^{2}$ ($\hat{g}_{\mu\nu}$ is the maximally symmetric metric)} in critical gravity, the Einstein field equations are of second order at the critical point. So it is possible to get analytical solutions in this higher-derivative  gravity theory and to give some insight into some interesting questions. In Ref. \cite{Liu:2012mia}, flat brane scenario in five-dimensional critical gravity was investigated and some analytical solutions were found for thin brane with the use of the junction conditions and for thick brane. It was found that scalar perturbations for all these brane solutions are stable.

However, flat brane scenario is the simplest case for the study of brane world. It is known that there are three types of branes with maximally symmetry: flat, de Sitter, and anti-de Sitter branes. Furthermore, the study of Friedmann每Robertson每Walker (FRW) brane is also interesting. It is expected that the warped brane models in higher derivative gravity such as critical gravity may provide new scenario in the study of AdS/CFT correspondence, cosmologies, and phenomenological model buildings\cite{Nojiri:2001ae,Soda:2010si}. In this paper, we generalize the work of Ref. \cite{Liu:2012mia} and construct the flat and warped brane solutions in $n$-dimensional critical gravity.
The organization of this paper is as follows. In Sec. II, we first derive the Gibbons-Hawking boundary term on the thin brane and give the the junction conditions. Then we construct analytic flat and warped brane solutions with the junction conditions and investigate the hierarchy problem in the thin brane scenario. In Sec. III, thick branes generated by a scalar field are investigated and the
conditions of the splitting of the branes are obtained. Finally, our conclusion
is given in Sec. IV.

\section{Thin brane solutions}
First we consider the thin brane model in the frame of a $n$-dimensional
critical gravity. The action is \cite{Deser:2011xc}
\begin{eqnarray}
    S = \frac{1}{2\kappa^{2}}\int d^{n}x \sqrt{-g} \mathcal{L}_{\mathrm{G}} + S_{\mathrm{B}},
    \label{action}
\end{eqnarray}
where
\begin{eqnarray}
    \mathcal{L}_{\mathrm{G}}= R - (n-2) \Lambda_{0} + \alpha R^{2}
        + \beta R_{MN}R^{MN} +
        \gamma \mathcal{L}_{\mathrm{GB}}  ,  \label{LG1}
\end{eqnarray}
and $\kappa$ denotes the $n$-dimensional gravitational
constant with $\kappa^{2}=8\pi/M_{*}^{4}$, where $M_{*}$ is the $n$-dimensional Planck mass
scale. The parameters $\alpha$ and $\beta$ satisfy the following critical condition \cite{Deser:2011xc}
\begin{eqnarray}
    {4(n-1)\alpha+n\beta}=0.  \label{CriticalCondition}
\end{eqnarray}
The brane part $S_{\mathrm{B}}$ of the above action is given by
\begin{equation}
    S_{\mathrm{B}}=\int d^{n-1}x\sqrt{-q}(-V_{0}),
\end{equation}
where $V_0$ is the brane tension and $q_{\mu\nu}$ is the induced metric on
the brane, which is assumed located at the origin of the extra dimension $x^n=y$.
The capitals letters $M,N,\cdots=0,1,2,3,\cdots,n-2,n$ and the Greek letters
$\mu,\nu,\cdots=0,1,2,\cdots,n-2$ denote the indices of the $n$-dimensional
bulk and $(n-1)$-dimensional brane, respectively.

The equations of motion (EoMs) derived from the action (\ref{action}) are read as
\begin{eqnarray}
    \mathcal{G}_{MN}+E_{MN}-\frac{\gamma}{2}H_{MN} = \kappa^{2}T_{MN},
    \label{EoM}
\end{eqnarray}
where
\begin{eqnarray}
     \mathcal{G}_{MN} &=&  R_{MN}-\frac{1}{2} R~ g_{MN}
         + \frac{1}{2}(n-2) \Lambda_0 g_{MN}, \label{gMN of EoM} \\
     E_{{MN}} &=&  2 \alpha R \left(R_{{MN}}
           -\frac{1}{4}R ~g_{{MN}}\right) \nonumber \\
          && + (2\alpha + \beta ) ( g_{{MN}} \square
           - \nabla_M\nabla_N ) R \nonumber \\
          &&+ 2 \beta R^{{PQ}} \left(R_{{MPNQ}}
           - \frac{1}{4} R_{{PQ}}~g_{{MN}}\right)\nonumber \\
           &&+\beta\square \left(R_{{MN} }-\frac{1}{2} R~ g_{MN}\right),  \label{EMN of EoM}  \\
    H_{MN}&=& g_{MN}\mathcal{L}_{\mathrm{GB}}
           -4 R R_{MN} + 8 R_{MP} R^{P}_{N} \nonumber\\
           &&
           + 8 R_{MANB} R^{AB} - 4R_{MABC} R_{N}^{~~ABC}
            \label{HMN of EoM}, \\
    T_{MN}&=& - \frac{2}{\sqrt{-g}} \frac{\delta S_{B}}{\delta g^{MN}}.
    \label{energy-momentum}
\end{eqnarray}

In $n$-dimensional space, we have the following relation
\begin{eqnarray}
  && \alpha R^2+\beta R_{MN}R^{MN}+\gamma\mathcal{L}_{\mathrm{GB}} \nonumber \\
  &&=  \frac{(n-2)\beta}{4(n-3)}C^2
       -\frac{\zeta}{4(n-3)}\mathcal{L}_{\mathrm{GB}} \nonumber \\
     &&  +\frac{4(n-1)\alpha+n\beta}{4(n-1)}R^2, \label{L2}
\end{eqnarray}
where
\begin{equation}
\zeta=(n-2)\beta-4(n-3)\gamma, \label{zeta}
\end{equation}
and $C^2:=C^{MNPQ}C_{MNPQ}$ is the square of the $n$-dimensional Weyl tensor,
\begin{eqnarray}
  C_{MNPQ}&=&R_{MNPQ}-\frac{2}{n-2}(g_{M[P}R_{Q]N}-g_{N[P}R_{Q]M}) \nonumber \\
          && + \frac{2}{(n-1)(n-2)}Rg_{M[P}g_{Q]N}.
\end{eqnarray}
Note that, under the critical
condition (\ref{CriticalCondition}), the last term $R^2$ in the right hand side of Eq. (\ref{L2})
vanishes.
So, the Lagrangian density $\mathcal{L}_G$ for the critical gravity can be reexpressed as
\begin{equation}
\mathcal{L}_{\mathrm{G}}
    = \mathcal{L}_{\mathrm{EGB}}+\frac{(n-2)\beta}{4(n-3)}C^2, \label{LG2}
\end{equation}
where $\mathcal{L}_{\mathrm{EGB}}$ is the Einstein-Gauss-Bonnet (EGB) term:
\begin{equation}
\mathcal{L}_{\mathrm{EGB}}
    = R - (n-2) \Lambda_{0}
       -\frac{\zeta}{4(n-3)}\mathcal{L}_{\mathrm{GB}}, \label{LEGB}
\end{equation}
%In the cases of flat, AdS, and dS branes, the Weyl tensor vanishes.

In the following, we first generalize the result of the junction conditions in five dimensions in Ref. \cite{Liu:2012mia} to $n$ dimensions for flat, AdS, and dS thin branes. Then we will use the generalized junction conditions to give the thin brane solutions.

\subsection{Junction conditions}

Following Ref. \cite{Liu:2012mia}, we adopt the Gibbons-Hawking method to derive the junction conditions. The basic idea is as follows. The whole spacetime $M$ is divided into two submanifolds by the thin brane, which is the boundary
$\partial M$ of the two submanifolds. The unit vector normal to the boundary $\partial M$ is denoted by $n^Q$ and it is outward pointing. Then the induced metric on the brane is $q^{MN}=g^{MN}-n^M n^N$. The
extrinsic curvature is defined as $K_{MN}=\mathcal{L}_{\vec{n}} q_{MN}/2 $. We denote  $[F]_{\pm}:=F(0+)-F(0-)$. In the following, we let
$n^{Q}(0_{+})=n^{Q}:=(0,0,0,0,-1)$ and $n^{Q}(0_{-}):=(0,0,0,0,+1)$ for the right and left sides, respectively. Due to the $Z_2$ symmetry of the extra dimension, we only need to calculate the right side. See e.g. Refs. \cite{Parry:2005eb,Balcerzak:2007da,Liu:2012mia} for the details.

We will deal with the $C^2$ term and EGB term, respectively.
We first consider a general geometry instead of the special case of branes with $ds^2=e^{2A(y)}\hat g_{\mu\nu}(x)dx^\mu dx^\nu\\
+dy^2$. For the $C^2$ term, we have
\begin{eqnarray}
 &&\delta \!\! \int_M \!\!\!\!\!\! d^{n} x \sqrt{-g}  C^2 %\nonumber \\
%  &=& \int_M d^5x\sqrt{-g}\bigg(
%        2C_{M}^{\;\;\;PQR}C_{NPQR}
%       -\frac{1}{2}g_{MN}C^2  \nonumber\\
%  &+& \frac{8}{3}R^{PQ}C_{MPNQ}-4C^{~~~P~~Q}_{(M~N)~;PQ}\bigg)\delta g^{MN}\nonumber\\
%  &+&  4\int_{\partial M} d^{4}x \sqrt{-q}
%      \Big(C^{MPNQ}n_Q \delta g_{MN;P}-C^{MPNQ}_{~~~~~~;Q}n_P \delta g_{MN} \Big). \label{CEoM}\\
  \!\!\!\supset \!\!
  %\int_M d^5x\sqrt{-g} \Big[2C_{M}^{~PQR}C_{NPQR}-\frac{1}{2}g_{MN}C^2 \nonumber \\
%  &+& \frac{8}{3}R^{PQ}C_{MPNQ}-4C^{~~~P~~Q}_{(M~N)~;PQ}\Big]
%       \delta g^{MN}\nonumber\\
     4\int_{\partial M} d^{n-1} x \sqrt{-g}  \Big[(C^{MPNQ}n_Q \delta g_{MN})_{;P} \nonumber \\
  && \!\!\!\! - \!\! \left(\!\!(C^{MPNQ}n_Q)_{;P}+C^{MPNQ}_{~~~~~~;Q}n_P \right) \! \delta g_{MN}\!\Big]. \label{Cboundary2}
\end{eqnarray}
Here, the bulk term has been omitted and only the relevant boundary term is given explicitly.
In order to have a well-posed variational principal, we introduce an auxiliary
field $\varphi^{MNPQ}$ and replace $C^2$ with $2\varphi^{MNPQ}C_{MNPQ}\\
-\varphi^{MNPQ}\varphi_{MNPQ}$.
Then from the EoM of the auxiliary field, $\varphi^{MNPQ}=C^{MNPQ}$, we can see that $\varphi^{MNPQ}$ has
the same symmetry as the Weyl tensor and is also totally traceless. With the new field
$\varphi^{MNPQ}$, Eq. (\ref{Cboundary2}) becomes
\begin{eqnarray}
 && \delta \!\! \int_M \!\!\!\! d^{n} x \sqrt{-g}  C^2 \!\!\!
  \supset
     4\int_{\partial M} d^{n-1} x \sqrt{-g}  \Big[(\varphi^{MPNQ}n_Q \delta g_{MN})_{;P} \nonumber \\
  &&\!\!\!\! -  \left( \! (\varphi^{MPNQ}n_Q)_{;P}+\varphi^{MPNQ}_{~~~~~~;Q}n_P \! \right) \! \delta g_{MN}\Big]. \label{Cboundary3}
\end{eqnarray}

Then with the identity \cite{Liu:2012mia}
 \begin{eqnarray}
 % \delta n_M&=&-\frac{1}{2}n_M n_P n_Q\delta g^{PQ},\label{basic1}\\
X^M_{;M}&=&D_M(q^{M}_N X^N)+ K n_NX^N+\mathcal{L}_{\vec{n}}(n_NX^N),\label{basic2}
 \end{eqnarray}
where $D_M(q^P_N X^N):=q_M^Q q^P_R(q^R_N X^N)_{;Q}$, we can show that
\begin{eqnarray}
  &&\int_{\partial M} d^{n-1} x \sqrt{-g} (\varphi^{MPNQ}n_Q \delta g_{MN})_{;P}\nonumber\\
%  &=&\int_{\partial M} [K\varphi^{MPNQ}n_Q n_P \delta g_{MN}+\mathcal{L}_{\vec{n}}(\varphi^{MPNQ}n_Q n_P \delta g_{MN})]\nonumber\\
 &=& \int_{\partial M} d^{n-1} x \sqrt{-g} \Big[
     \varphi^{MPNQ}n_Q n_P\mathcal{L}_{\vec{n}} \delta g_{MN}  \nonumber\\
  &&  \!\!\!\! + \Big(K\varphi^{MPNQ}n_Q n_P
          +\mathcal{L}_{\vec{n}}(\varphi^{MPNQ}n_Q n_P)
     \Big) \delta g_{MN}
    \Big]. \label{phinndg}
\end{eqnarray}
Further, the first term in the above equation can be reduced to
\begin{eqnarray}
 &&  \varphi^{MPNQ}n_Q n_P\mathcal{L}_{\vec{n}} \delta g_{MN} \nonumber\\
 &=& 2\varphi^{MN}\delta K_{MN}
     -\varphi^{MN}K_{MN}n^P n^Q\delta g_{PQ}\nonumber\\
 &&  +2D_N(\varphi^{MN}n^P\delta g_{PM}) -2\varphi^{PM}_{;P}n^N\delta g_{MN},
\end{eqnarray}
where $\varphi^{MN}:=\varphi^{MPNQ}n_Q n_P$.
So the surface term for the $C^2$ part is
\begin{eqnarray}
  \delta S_{C^2} &\supset&
   \delta \left(\frac{1}{2\kappa^2} \frac{(n-2)\beta}{4(n-3)}
   \int_M d^{n-1} x \sqrt{-g}  {C^2}\right) \nonumber \\
 &=& \frac{(n-2)\beta}{2(n-3)\kappa^2}\int_{\partial M} d^{n-1} x \sqrt{-g}
      \Big\{ 2\varphi^{MN}\delta K_{MN}
      \nonumber\\
     && +\Big[
        \mathcal{L}_{\vec{n}}\varphi^{MN}
        + K\varphi^{MN}
        - \varphi^{PQ}K_{PQ}n^M n^N      \nonumber\\
     &&
        - 2\varphi^{P(M}_{;P}n^{N)}   - (\varphi^{MPNQ}n_Q)_{;P}
       \nonumber\\
     &&
        - \varphi^{MPNQ}_{~~~~~~;Q}n_P \Big] \delta g_{MN}\Big\}. \label{c-bound-g1}
\end{eqnarray}
Then with $2\varphi^{MN}\delta K_{MN} = 2\varphi^{MN}\delta (K_{MN}-\frac{1}{n-1}q_{MN}K)+\frac{2}{n-1}K\varphi^{MN}\delta q_{MN}$, we finally obtain
\begin{eqnarray}
  \delta S_{C^2}
     &\supset& \frac{(n-2)\beta}{2(n-3)\kappa^2}\int_{\partial M} d^{n-1} x \sqrt{-g}
      \Big[ 2\varphi^{MN}\delta\bar{K}_{MN}  \nonumber\\
     &&    +\big(W^{MN}-\varphi^{PQ}K_{PQ}n^M n^N \big) \delta g_{MN}
      \Big],  \label{c-bound-g}
 \end{eqnarray}
 where
 \begin{eqnarray}
 \bar{K}_{MN}&:=&K_{MN}-\frac{1}{n-1}q_{MN}K, \\
   W^{MN}&:=&\frac{n+1}{n-1}K\varphi^{MN}+\mathcal{L}_{\vec{n}}\varphi^{MN}
       -2\varphi^{P(M}_{;P}n^{N)}  \nonumber \\
    && {-(\varphi^{MPNQ}n_Q)_{;P}
       -\varphi^{MPNQ}_{~~~~~~;Q}n_P}.
 \end{eqnarray}
It can be shown that $W^{MN}n_M=W^{MN}q_{MN}=$

$\!\!\!\!\!\!\!\!W^{MN}g_{MN}=0$.

Now we can introduce the corresponding Gibbons-Hawking surface term \cite{Hawking:NP1984} for the $C^2$ term
\begin{eqnarray}
  S_{C^2\mathrm{-suf}}=-\frac{(n-2)\beta}{(n-3)\kappa^2}\int_{\partial M} d^{n-1} x \sqrt{-g}~\varphi^{MN} \bar{K}_{MN}.
  \end{eqnarray}
So we have (considering the whole spacetime)
%\footnote{The terms $q^{MN}\bar{K}^{PQ}\bar{K}_{PQ}$, $\bar{K}^{MP}\bar{K}^{N}_P$, $\tilde{R}^{MN}-\frac{1}{4}q^{MN}\tilde{R}$ have no contribution to junction conditions since Eq. (\ref{barK}) .}
\begin{eqnarray}
 &&\delta (S_{C^2}+S_{C^2\mathrm{-suf}}) %\nonumber \\
  = \frac{(n-2)\beta}{2(n-3)\kappa^2}\int_{\partial M} d^{n-1} x \sqrt{-g} \nonumber \\
   &&    \Big\{-2 \big[\bar{K}_{MN}\big]_{\pm} {\delta \varphi^{MN}}  -  \big[\varphi^{PQ}K_{PQ}\big]_\pm n^M n^N\delta g_{MN} \nonumber \\
    &&  +\big[W^{MN}\big]_\pm\delta g_{MN} \Big\}. \label{deltaS1}
\end{eqnarray}

Next, we come to the EGB term in Eq. (\ref{LEGB}), for which the Gibbons-Hawking surface term was given in Refs. \cite{Deruelle:2000ge,Davis:2002gn,Maeda:2003vq}:
\begin{eqnarray}
S_{\mathrm{EGB-surf}} \!&=&
\! \frac{1}{2\kappa^2}\int_{\partial M} \!\!\!\!\!\! d^{n-1} x \sqrt{-g} \nonumber \\
&& \times \Big(2K-\frac{\zeta}{(n-3)}(J-2\tilde{G}_{\mu\nu}K^{\mu\nu})\Big)
    \end{eqnarray}
with $\tilde{G}_{\mu\nu}=\tilde{R}_{\mu\nu}-q_{\mu\nu}\tilde{R}/2$ the Einstein tensor of the induced metric $q_{\mu\nu}$ and $J$ the trace of the following tensor:
 \begin{eqnarray}
J_{MN}&=&\frac{1}{3}\Big(2K K_M^P K_{PN}+K^{PQ}K_{PQ}K_{MN} \nonumber \\
   &&-K^2K_{MN}-2K_{MP}K^{PQ}K_{QN}\Big).
 \end{eqnarray}
Then we have
\begin{eqnarray}
   && \delta (S_{\mathrm{EGB}}+S_{\mathrm{EGB-surf}}) \nonumber \\
   &=&  \frac{1}{2\kappa^2}
     \delta \int_M d^{n}x \sqrt{-g}
           \Big[ R - (n-2) \Lambda_{0} -\frac{\zeta}{4(n-3)}\mathcal{L}_{\mathrm{GB}}  \Big]  \nonumber \\
   &+&  \frac{1}{2\kappa^2}
       \delta \int_{\partial M} d^{n-1} x \sqrt{-g}
      \Big(2K-\frac{\zeta}{(n-3)}(J-2\tilde{G}_{\mu\nu}K^{\mu\nu})\Big) \nonumber\\
  &\supset& \frac{1}{2\kappa^2}\int_{\partial M} d^{n-1} x \sqrt{-g}
       (-E_{\mathrm{EGB}}^{MN})\delta g_{MN}, \label{deltaS2}
\end{eqnarray}
where
\begin{eqnarray}
 E_{\mathrm{EGB}}^{MN}&:=& [K^{MN}]_{\pm}-q^{MN}[K]_{\pm} \nonumber \\
      &&-\frac{\zeta}{2(n-3)}
      \Big(3[J^{MN}]_{\pm}
       -q^{MN} [J]_{\pm} \nonumber \\
     &&  -2P^{MPNQ}[K_{PQ}]_{\pm}\Big), \label{EGB JunctionCondition} \\
P_{MNPQ}
     &:=& \tilde{R}_{MNPQ}
      -2q_{M[Q}\tilde{R}_{P]N} \nonumber \\
      && +2q_{N[Q}\tilde{R}_{P]M}
      +\tilde{R}q_{M[P}q_{Q]N}.
 \end{eqnarray}

Thus, from Eqs. (\ref{deltaS1}) and (\ref{deltaS2}), for $n$-dimensional critical gravity theory,
we finally get
\begin{eqnarray}
 &&\delta \big( S_{\mathrm{EGB}}+S_{C^2}
   +S_{\mathrm{EGB-surf}}+S_{C^2\mathrm{-suf}} \big) \nonumber \\
  &\supset& \frac{1}{2\kappa^2}\int_{\partial M} d^{n-1} x \sqrt{-g}
       \Big\{\frac{(n-2)\beta}{(n-3)}\Big(-2 \big[\bar{K}_{MN}\big]_{\pm} {\delta \varphi^{MN}}  \nonumber \\
  &-&  \big[K_{PQ}\varphi^{PQ}\big]_\pm n^M n^N\delta g_{MN}
      +\big[W^{MN}\big]_\pm\delta g_{MN} \Big)   \nonumber \\
   &-&   E_{\mathrm{EGB}}^{MN}\delta g_{MN}  \Big\}. \label{deltaS}
\end{eqnarray}
So, the junction conditions are
 \begin{eqnarray}
\big[\bar{K}_{MN}\big]_\pm &=&0,\label{barK}\\
\big[K_{PQ}\varphi^{PQ}\big]_\pm=\bar{K}_{PQ}[\varphi^{PQ}]_\pm&=&0,\\
 \big[E^{MN}_{\mathrm{EGB}}\big]_\pm
    -\frac{(n-2)\beta}{n-3}\big[W^{MN}\big]_\pm
    &=&\kappa^2T_{(\mathrm{brane})}^{MN}\label{c-jun}.
 \end{eqnarray}
Here $T_{(\mathrm{brane})}^{MN}$ denotes the singular part of $T^{MN}$.
To avoid the $\delta$-function in the junction conditions, we need the stronger condition $[\varphi^{MN}]_\pm=0$.
%For this case, Eq. (\ref{c-jun}) becomes
%\begin{eqnarray}
%-\frac{(n-2)\beta}{n-3}\big[W^{MN}\big]_\pm+\big[E_{\mathrm{GB}}^{MN}\big]_\pm =-\kappa^2T_{(\mathrm{brane})}^{MN}\label{c-jun1}.
% \end{eqnarray}

For the special warped geometries of flat, AdS, and dS branes, whose metrics have the form
 \begin{eqnarray}
ds^2=e^{2A(y)}\hat g_{\mu\nu}(x)dx^\mu dx^\nu+dy^2,
 \end{eqnarray}
the first condition (\ref{barK}) gives no more constraint for brane solutions because $\bar{K}_{MN}\equiv0$; and
$C^{MPNQ}$ is continuous and its contribution vanishes. So the above junction conditions for flat, AdS, and dS brane solutions in the critical gravity are simplified as
 \begin{eqnarray}
\big[E^{MN}_{\mathrm{EGB}}\big]_\pm
    =\kappa^2T_{(\mathrm{brane})}^{MN}\label{c-jun},
 \end{eqnarray}
where the nonvanishing components the brane energy-momentum tensor are $T_{\mu\nu}^{\mathrm{brane}}=-V_0 \hat{g}_{\mu\nu}$. The reduced metric is $q_{\mu\nu}=\hat{g}_{\mu\nu}(x)e^{2A(y)}$. With the constraint $A(0)=0$ and the assumption of the $Z_2$ symmetry of the extra dimension ($A(y)=A(-y)$), we have $K_{\mu\nu}(0_+)=-K_{\mu\nu}(0_-)=-A'(0_+)\hat{g}_{\mu\nu}$, and hence $[K_{\mu\nu}]_{\pm}-q_{\mu\nu}[K]_{\pm}=2(n-3)A'(0_{+})\hat{g}_{\mu\nu}$.
%The explicit junction conditions will be given in the following subsections for the branes with maximum symmetry, namely, Minkowski (flat), AdS, and dS branes.

Next, we mainly consider the branes with maximum symmetry, namely, flat (Minkowski),
AdS, and dS branes. With the explicit junction conditions, we will give the thin brane solutions. The flat brane solutions in five-dimensional critical gravity
has been found in Ref.~\cite{Liu:2012mia}.

\subsection{Flat brane}

The line-element of a flat brane with the most
general $(n-1)$-dimensional Poincar\'{e}-invariant is
\begin{equation}
ds^2 = \mathrm{e}^{2A(y)} \eta_{\mu\nu} dx^{\mu} dx^{\nu} + dy^{2},
    \label{flat metric}
\end{equation}
where $ \mathrm{e}^{2A(y)} $ is the warp factor. Such a compactification is known as a warped compactification. Considering the $ Z_2 $ symmetry
of the brane model, we have $ A(y)=A(-y)$. Furthermore, we can set $ \mathrm{e}^{2A(0)}=1$
in order to get $q^{\mu\nu}=\eta^{\mu\nu}$ on the brane. The bulk energy-momentum tensor reads
\begin{equation}
    T_{MN}=-V_0 \delta^{\mu}_{M} \delta^{\nu}_{N}\mathrm{e}^{2A(y)} \eta_{\mu\nu} \delta(y),
    \label{TMN}
\end{equation}
from which the brane energy-momentum tensor is given by
\begin{equation}
    T_{MN}^{\mathrm{brane}}=-V_0 \delta^{\mu}_{M} \delta^{\nu}_{N} \eta_{\mu\nu}
    ~~~\mathrm{or} ~~~
    T_{\mu\nu}^{\mathrm{brane}}=-V_0 \eta_{\mu\nu}.
    \label{TMN}
\end{equation}

For arbitrary $ \alpha $ and $ \beta $, the field equations (\ref{EoM})
are forth-order differential ones. However, at the critical
point $\alpha = - \frac{n}{4(n-1)} \beta$ \cite{Deser:2011xc}, the bulk field equations turn out to be
\begin{eqnarray}
   [2+(n-4)\zeta A'^{2}]A''&=&0  \label{thin flat EoM 1},\\
    4 \Lambda_{0} + (n-1) A'^{2} [4 + (n-4) \zeta A'^{2}]&=&0, \label{thin flat EoM 2}
\end{eqnarray}
where $\zeta$ is given by Eq. (\ref{zeta}), and  the prime and double prime stands
for the first-order and second-order derivations with respect to $y$, respectively.
Throughout this paper we will use the critical condition (\ref{CriticalCondition}).
The junction conditions read
%\begin{eqnarray}
%    \int_{0^{-}}^{0^{+}}dy [2+(n-4)\zeta A'^{2}]A''
%    =\Big({2A'+\frac{1}{3}(n-4)\zeta A'^3}\Big)\Big|_{0^-}^{0^+}
%    =-\frac{2\kappa^{2}}{n-2}V_{0}.
%\end{eqnarray}
\begin{eqnarray}
    (n-2)\Big[{A'+\frac{1}{6}(n-4)\zeta A'^3}\Big]_{\pm}
    =-\kappa^{2}V_{0},
\end{eqnarray}
or
\begin{eqnarray}
    2(n-2)\Big({A'(0_{+})+\frac{1}{6}(n-4)\zeta A'^3(0_{+})}\Big)
    =-\kappa^{2}V_{0},\label{JunctionConditionFlat}
\end{eqnarray}
due to the $Z_2$ symmetry of the extra dimension.

The solution for the warp factor $A(y)$ is
\begin{eqnarray}
    A(y) \!&=&\! -k |y|,   \label{Ay2}
\end{eqnarray}
where $k$ is a positive parameter since we are interested in the exponentially decreasing warp factor, which could solve the hierarchy problem if we consider the two-brane model with an $S^1/Z_2$ extra dimension \cite{Randall:1999ee}.
Then, from Eq. (\ref{thin flat EoM 2}), the naked cosmological constant is given by
\begin{eqnarray}
    \Lambda_0 = - (n-1) \left(1 + \frac{1}{4}(n-4) \zeta k^2 \right) k^2.
\end{eqnarray}
The brane tension is determined by the junction condition (\ref{JunctionConditionFlat}):
\begin{eqnarray}
    V_0=\frac{n-2}{3\kappa^{2}}
        \left(6+(n-4)\zeta k^2\right)  k.  \label{V02}
\end{eqnarray}
It is clearly that the result is consistent with the one in general relativity when $\zeta=0$. The naked cosmological constant and brane tension are respectively negative and positive when $\zeta=0$ (in this paper, we assume that $n\geq 5$), and can be positive, zero, and negative when $\zeta<0$, depending on the magnitude of $\zeta$ compared with $k^{-2}$.
If we require that the higher-order terms in (\ref{LG1}) are small compared with the $R$ term, which implies $\zeta k^2 \ll 1$, then we will have negative $\Lambda_0$ and positive brane tension for any such $\zeta$.
If we rewrite the Einstein equations (\ref{EoM}) as $G_{MN}=\kappa^{2}T^{\mathrm{eff}}_{MN}$, namely, and identify
$(E_{MN}-\frac{\gamma}{2}H_{MN})/\kappa^2 + T_{MN}$ as an effective energy-momentum tensor,
then we will always get an effective positive brane tension.

The flat thin brane is embedded in an AdS$_{n}$
spacetime, with the effective cosmological constant $\Lambda$ given by $\Lambda = -(n-1)k^2$. Therefore, the higher-order terms only effect the naked cosmological constant and brane tension.

For the case $n=5$, the result reads
\begin{eqnarray}
    \Lambda_0 &=&
     -(4 +  \zeta k^2)k^2,   \\
 V_{0} &=&
        {\kappa^{-2}}\left(6+\zeta k^2\right)  k,
\end{eqnarray}
which is the thin brane solution found in Ref.~\cite{Liu:2012mia}.

\subsection{AdS brane}
The metric describing an AdS brane embedded in an AdS$_n$ spacetime is assumed as
\begin{eqnarray}
    ds^2 &=& e^{2A(y)} \big[\mathrm{e}^{2Hx_{n-2}} (-dt^{2} + dx_{1}^{2} +\cdots+ dx_{n-3}^{2} )
      \nonumber \\
    &&+ dx_{n-2}^{2} \big]+ dy^{2}.
    \label{AdS metric}
\end{eqnarray}
The corresponding Einstein equations beyond the thin brane turn out to be
\begin{eqnarray}
   && \!\!\!\! (\mathrm{e}^{2A} A'' -H^{2})\!\!
    \left[ 2 + (n-4 )H^{2}\zeta\mathrm{e}^{-2A} \!+\!(n-4)\zeta A'^{2} \right]
    = 0, \label{EoMAdSbrane1} \nonumber \\
    ~\\
   && \!\!\!\! (n-1) A'^{2}
    \left[ 2(n-4)H^{2}\zeta  + \mathrm{e}^{2A}(4+(n-4)\zeta A'^{2}) \right]
      \nonumber  \\
    && \!\!\!\!+4(n-1) H^{2}
     +  (n-1)(n-4)H^{4}\zeta\mathrm{e}^{-2A}
     + 4\Lambda_{0}\mathrm{e}^{2A}  =0. \nonumber \\ \label{EoMAdSbrane2}
\end{eqnarray}
The junction conditions read
\begin{eqnarray}
    (n-2)\Big[A'+\frac{1}{6}(n-4)\zeta (A'^3+3H^{2}A')\Big]_{\pm}
    = -{\kappa^{2}}V_{0}.﹛﹛\label{JunctionConditionAdS}
\end{eqnarray}
The solution for Eq.~(\ref{EoMAdSbrane1}) is
\begin{eqnarray}
    A(y) \!=\! \ln \left[ \frac{H}{k} \cosh(k|y| + \sigma) \right] \label{AdSbrane_Ay}
\end{eqnarray}
with
\begin{eqnarray}
    \sigma &=& \mathrm{arccosh}(\frac{k}{H}).
\end{eqnarray}
Here, the two parameters should satisfy the relation: $k>H$.
Substituting the solution (\ref{AdSbrane_Ay}) into the second equation
(\ref{EoMAdSbrane2}), we get the naked cosmological constant:
\begin{eqnarray}
    \Lambda_0 = - (n-1)  \left(1+ \frac{1}{4}(n-4)\zeta k^2  \right)k^2.
\end{eqnarray}
Since $R_{MN}=\Lambda g_{MN}=-(n-1)k^2 g_{MN}$, the cosmological constant
of the AdS$_{n}$ is $\Lambda =-(n-1)k^2$.
The naked cosmological constant can be positive, zero, and negative, depending
on the value of the combine of the parameters $\beta$ and $\gamma$ (i.e., $\zeta=-4\gamma(n-3)+\beta(n-2)$).
The junction conditions (\ref{JunctionConditionAdS}) give  the brane tension:
\begin{eqnarray}
    ~V_0 = \frac{n-2}{3\kappa^2} \left[ 6+ (n-4)(k^2+2H^2)  \zeta \right] \sqrt{k^2-H^2},
\end{eqnarray}
which can be positive, negative, or zero.
%when the parameter $\zeta$ is greater than, less than, and equals $<-6/[(n-4)(k^2+2H^2) ]$,$\zeta>-6/[(n-4)(k^2+2H^2) ]$

\subsection{dS brane}
The metric describing a dS brane has the following form:
\begin{equation}
    ds^2 = \mathrm{e}^{2A(y)} [dt^{2}+\mathrm{e}^{-2Ht}\delta_{ij} dx^{i} dx^{j}]+ dy^{2},
    \label{dS metric}
\end{equation}
The EoMs at $y\neq0$ are
\begin{eqnarray}
  && \!\!\!\! (\mathrm{e}^{2A} A''\!+\!H^{2} )\!\!\left[ 2
       -(n-4) H^{2}\zeta  \mathrm{e}^{-2A}
       \! +\! (n-4)\zeta   A'^{2} \right]
    \!\!=\!\!0, \nonumber \\~\\
  &&\!\!\!\!  (n-1) A'^{2}
    \left[ -2(n-4)H^{2}\zeta  + \mathrm{e}^{2A}(4+(n-4)\zeta A'^{2}) \right]
      \nonumber\\
  &&\!\!\!\!  -4(n-1) H^{2}
     +  (n-1)(n-4)H^{4}\zeta\mathrm{e}^{-2A}\!
     + \! 4\Lambda_{0}\mathrm{e}^{2A} \!\! =\!\!0. \nonumber\\
    \label{thin dS EoM}
\end{eqnarray}
The junction condition is similar with the case of AdS brane:
%\begin{eqnarray}
%    &&\int_{0^{-}}^{0^{+}}dy \left[2-(n-4)H^{2}\zeta +(n-4)\zeta A'^{2} \right]
%    (A''+H^{2}) \nonumber \\
%    %&=&\int_{0^{-}}^{0^{+}}dy \left[-H^{2}(n-4)\zeta A'^{2}
%%    + (2+H^{2}\zeta (n-4))A''
%%    + (n-4)\zeta A'^{2}A'' \right] \nonumber \\
%    &=& \Big({(2-(n-4)H^{2}\zeta )A'
%               +\frac{1}{3}(n-4)\zeta A'^3}\Big)\Big|_{0^-}^{0^+} \nonumber \\
%    &=& -\frac{2\kappa^{2}}{n-2}V_{0}.﹛﹛\label{JunctionConditionAdS}
%\end{eqnarray}
\begin{eqnarray}
    (n-2)\Big[A'+\frac{1}{6}(n-4)\zeta (A'^3-3H^{2}A')\Big]_{\pm}
    = -{\kappa^{2}}V_{0}.﹛﹛\label{JunctionConditiondS}
\end{eqnarray}
The solution is
\begin{eqnarray}
    A(y) \!=\! \ln \left[ \frac{H}{k} \sinh(k|y| + \sigma) \right],
\end{eqnarray}
where
\begin{eqnarray}
    \sigma &=& \mathrm{arcsinh}(\frac{k}{H}).
\end{eqnarray}
The naked cosmological constant and other parameters are related by
%\begin{eqnarray}
%    \Lambda_0 = -\frac{1}{4} k^2 (n-1) \left(k^2 n \zeta -4 k^2 \zeta +4\right),
%\end{eqnarray}
\begin{eqnarray}
    \Lambda_0 = - (n-1) k^2 \left(1+ \frac{1}{4}(n-4) k^2 \zeta \right).
\end{eqnarray}
%The cosmological constant of the AdS$_{n}$ is also $\Lambda =-(n-1)k^2$ and
%the relation between $\Lambda$ and $\Lambda_0$ is (\ref{RelationBetween2Lambdas}).

The junction condition gives the relation between the brane tension and other parameters:
\begin{eqnarray}
    V_0 = \frac{n-2}{3\kappa^2}\left[ 6+ (n-4)(k^2-2H^2) \zeta  \right] \sqrt{k^2+H^2}.
\end{eqnarray}
Just as the case of AdS brane, the brane tension here can also be positive, negative, or zero.

{For the flat, AdS, and dS thin branes, the metric at $y\neq0$ satisfies $ R_{MN}=\Lambda g_{MN}$, where $\Lambda$ is the $n$-dimensional effective cosmological constant. It is because that these thin branes are all embedded in AdS$_{n}$  spacetime. $\Lambda$ is related to the naked cosmological constant $\Lambda_{0}$ by
\begin{eqnarray}
    \Lambda_0 =\Lambda-\frac{n-4}{4(n-1)}\zeta \Lambda^{2}.
    \label{lambda}
\end{eqnarray}
As discussed in Ref.~\cite{Deser:2011xc}, Eq. (\ref{lambda}) has two roots, which
correspond to two AdS vacuums. One of the AdS vacuums has negative energy
excitations. On the other hand,
to render the massive spin-2 mode into massless, we need a second critical condition \cite{Deser:2011xc}
\begin{eqnarray}
    1+\frac{[-(n-2)^{2}\beta+4(n-3)(n-4)\gamma]\Lambda}{2(n-1)}=0.
    \label{critical2}
\end{eqnarray}
Under this condition, one obtains the effective cosmological constant $\Lambda_{\texttt{c}}$ of the critical vacuum
\begin{eqnarray}
   \Lambda_{\texttt{c}}=-\frac{n(n-1)}{2(n-1)(n-2)^{2}\alpha+2n(n-3)(n-4)\gamma},
\end{eqnarray}
and the corresponding critical naked cosmological constant
\begin{eqnarray}
   \Lambda_{0\texttt{c}}\!\!=\!\!-\frac{n^{2}(n-1)\!\left[ (n-1)(n-2)\alpha+\!(n-3)(n-4)\gamma \right]}{2(n-1)(n-2)^{2}\alpha+2n(n-3)(n-4)\gamma}.\nonumber \\
\end{eqnarray}
As shown in Ref. \cite{Deser:2011xc}, for the critical vacuum, the
massive spin-2 mode become massless, and the
 excitation energy of the massless graviton vanishes; while for the noncritical vacuum,
 the spectrum still contains both massive and massless modes, and their excitation energies are of opposite signs.

Furthermore, if we require $\zeta=0$,  Eq. (\ref{lambda}) renders $\Lambda_{0}=\Lambda$, there will be only one AdS vacuum. Equation (\ref{LG1})
now is reduced to
\begin{eqnarray}
    \mathcal{L}_{\mathrm{G}}= R - (n-2) \Lambda_{0}+\frac{(n-2)\beta}{4(n-3)}C^{2}
\end{eqnarray}
with $\Lambda_{0}=-\frac{n}{8\alpha}$.
}

\subsection{Effective action and mass hierarchy}

To derive the effective action of gravity on the brane, we follow the procedure in Ref. \cite{Randall:1999ee}.
The $n$-th dimension coordinate ranges from $-y_b$ to $y_b$ with the topological of $S^1/Z_2$, brane I locates
at $y=0$, and brane II locates at $y=y_b$.
We consider the massless gravitational fluctuations of the background metric (\ref{flat metric}):
\begin{eqnarray}
    ds^2 &=& \mathrm{e}^{2A(y)} \hat{g}_{\mu\nu}(x) dx^{\mu} dx^{\nu} + dy^{2} \nonumber \\
         &=& \mathrm{e}^{2A(y)} (\eta_{\mu\nu} + h_{\mu\nu}(x)) dx^{\mu} dx^{\nu} + dy^{2}.
    \label{perturbatied flat metric}
\end{eqnarray}
These massless gravitational fluctuations are the zero modes of the classical solution (\ref{flat metric}) and $h_{\mu\nu}(x)$ is the physical graviton of the four-dimensional effective theory.
With the help of solution (\ref{Ay2}), we have
\begin{eqnarray}
    \sqrt{-g} \mathcal{L}_{\mathrm{G}}
    &=&\sqrt{-g} \big[ R - (n-2) \Lambda_{0} + \alpha R^{2} \nonumber \\
    &&
        + \beta R_{MN}R^{MN} +
        \gamma \mathcal{L}_{\mathrm{GB}} \big] \nonumber \\
    &=&\sqrt{-\hat{g}} \Big[ a \mathrm{e}^{-(n-3)k|y|} \hat{R}\nonumber \\
    && + \mathrm{e}^{-(n-5)k|y|} \left( \alpha \hat{R}^{2}
        + \beta \hat{R}_{\mu\nu}\hat{R}^{\mu\nu} +
        \gamma \hat{\mathcal{L}}_{\mathrm{GB}} \right)  \nonumber \\
     && + ~\mathrm{function~of}~ A(y), ~A'(y), ~\mathrm{and}~ A''(y) \Big], \label{LG reduction}
\end{eqnarray}
where
\begin{eqnarray}
    a=1-2k^2 \left[-\frac{1}{4}(n-2)^2\beta+(n^2-5n+2)\gamma\right],
\end{eqnarray}
$M_{\texttt{pl}}$ is the $(n-1)$-dimensional Plank scale on the brane, $M$ is the $n$-dimensional Plank scale satisfying $2M^{n-2}=\frac{1}{2\kappa^2}$,
and terms like $\hat{g}$ and $\hat{R}$ are constructed by
$\hat{g}_{\mu\nu}(x)$. Note that in the action (\ref{LG1}) the terms like $R^2$ and $R_{MN}R^{MN}$ et al. are considered as higher-order terms comparing with the $R$ term. This is equivalent to  $\alpha R^2$, $\beta R_{MN}R^{MN}$, $\gamma {\mathcal{L}}_{\mathrm{GB}} {\ll} R$, which imply $\beta k^2,\gamma k^2 {\ll}1$. So we have $|1-a| \ll 1$.
The action of the $n$-dimensional gravity is reduced to
\begin{eqnarray}
    S_{\mathrm{G}} &=& 2M^{n-2}\int d^{n}x \sqrt{-g} \mathcal{L}_{\mathrm{G}}  \nonumber \\
                &\supset& 2aM^{n-2}\int_{-y_b}^{y_b} dy ~\mathrm{e}^{-(n-3)k|y|}
                    \int d^{n-1}x \sqrt{-\hat{g}}\hat{R}   \nonumber \\
                 && + 2M^{n-2}\int_{-y_b}^{y_b} dy ~\mathrm{e}^{-(n-5)k|y|} \nonumber \\
                 &&   \times\int d^{n-1}x \sqrt{-\hat{g}}\left( \alpha \hat{R}^{2}
                    + \beta \hat{R}_{\mu\nu}\hat{R}^{\mu\nu} +
                    \gamma \hat{\mathcal{L}}_{\mathrm{GB}} \right) \nonumber \\
                &\supset& S_{\mathrm{eff}},
    \label{SG reduction}
\end{eqnarray}
where the $(n-1)$-dimensional effective action is
\begin{eqnarray}
    S_{\mathrm{eff}}&=& 2 M_{\texttt{pl}}^{n-3} \int d^{n-1}x \sqrt{-\hat{g}} \nonumber \\
   && \times\left[\hat{R}
   + b\left( \alpha \hat{R}^{2}
                   + \beta \hat{R}_{\mu\nu}\hat{R}^{\mu\nu} +
                   \gamma \hat{\mathcal{L}}_{\mathrm{GB}} \right)\right],\\
   b&=& \left\{ \begin{array}{ll}
   \frac{(n-3)\left[ 1-\mathrm{e}^{-(n-5)ky_b} \right]}{a(n-5)\left[ 1-\mathrm{e}^{-(n-3)ky_b} \right]},
          &{ n\geq6},    \\
           \frac{(n-3)ky_b}{a\left[ 1-\mathrm{e}^{-(n-3)ky_b} \right]},
      &{ n=5}. \end{array} \right.
\end{eqnarray}
Here, the effective Planck scale $M_{\texttt{pl}}$ is related by the fundamental one $M$ via
\begin{eqnarray}
    M_{\texttt{pl}}^{n-3} &=&  \frac{2M^{n-3}}{(n-3)k} a\left[ 1-\mathrm{e}^{-(n-3)ky_b} \right].   \label{Mpl}
\end{eqnarray}
From this, we can see that the relationship between the fundamental scale and
the effective one reduces to the case in RS1 model because $a \doteq 1$.
Hence, the $n$-dimensional critical gravity reduces to the $(n-1)$-dimensional critical gravity on the brane.
Substituting the metric (\ref{perturbatied flat metric}) into the junction condition (\ref{c-jun}),
we obtain the brane tensions of the two branes:
\begin{eqnarray}
    V_I=-V_{II}=\frac{n-2}{3\kappa^{2}}
        \left(6+(n-4)\zeta k^2\right)  k.
\end{eqnarray}

Let us consider a Higgs field on the brane II with the action (for the case $n=5$)
\begin{eqnarray}
    S_H \!\! &=& \!\! \int \!\!\! d^{4}x \sqrt{-\hat{q}(y_b)} \nonumber \\
       &&     \times\left[ -\hat{q}^{\mu\nu}(y_b) D_\mu H^\dagger D_\nu H
            -\lambda (|H|^2-v_0^2)^2 \right],
\end{eqnarray}
where $v_0$ is the vacuum expectation value of the Higgs field.
%\begin{eqnarray}
%    S_H = \int d^{4}x \sqrt{-\hat{g}} \mathrm{e}^{-(n+1)ky_b}
%            \left[ \hat{g}^{\mu\nu} D_\mu H^\dagger D_\nu H
%            -\lambda (|H|^2-v_0^2)^2 \right].
%\end{eqnarray}
Redefining the field $\hat{H}=\mathrm{e}^{-ky_b}H$, we obtain the
canonical normalized action of the Higgs field $\hat{H}$:
\begin{eqnarray}
    S_H &=& \int d^{4}x \sqrt{-\hat{g}}  \nonumber \\
       &&   \times  \left[ -\hat{g}^{\mu\nu} D_\mu \hat{H}^\dagger D_\nu \hat{H}
            -\lambda (|\hat{H}|^2-\mathrm{e}^{-2ky_b}v_0^2)^2 \right].
\end{eqnarray}
Therefore, the vacuum expectation value of the Higgs field $\hat{H}$
would have a redshift due to the influence of the warped extra dimension
\begin{eqnarray}
    \hat{v}_0 = \mathrm{e}^{-ky_b}v_0,
\end{eqnarray}
which implies that the electro-weak scale has a redshift. On the other
hand, the mass of particles origins from the Yukawa coupling, and the vacuum
expectation value of the Higgs field is one of the parameters that determine
the mass. Hence, the effective (physical) mass also has a redshift
\begin{eqnarray}
    m = \mathrm{e}^{-ky_b}m_0.
\end{eqnarray}
From the above expression, we see that the redshift of the the vacuum expectation value of the Higgs field and the mass of the particles are the same with the RS1 model in Ref. \cite{Randall:1999ee}. So, the mass hierarchy problem is also solved in the higher-order braneworld model in the critical gravity.

\section{Thick branes generated by a scalar field}

In this section we study thick branes generated by a scalar field. The brane part of the action (\ref{action}) is
\begin{eqnarray}
    S_{\mathrm{B}}=\int d^{n}x\sqrt{-g}\big( -\frac{1}{2}g^{MN}\partial_{M}\phi\partial_{N}\phi-V(\phi) \big),
\end{eqnarray}
where the scalar field is assumed as $\phi=\phi(y)$ for flat, AdS, and dS branes considered below.

\subsection{Flat brane}

The line-element of a flat brane generated by a scalar field is also assumed
as (\ref{flat metric}). The EoMs (\ref{EoM}) reduce to the following second-order coupled equations:
\begin{eqnarray}
    &&   [2+(n-4)\zeta A'^{2}]A''=
          -\frac{2\kappa^{2}}{n-2}\phi'^{2} ,   \label{scalar flat Eom 1}   \\
   &&  4 \Lambda_{0} + (n-1) A'^{2} [4 + (n-4) \zeta A'^{2}]  \nonumber \\
    &&    =
        \frac{8\kappa^{2}}{n-2}
       \Big(\frac{1}{2}\phi'^{2}-V\Big),  \label{scalar flat Eom 2}  \\
   &&  \phi'' + (n-1)A'\phi'- V_{\phi} = 0,
    \label{scalar flat Eom 3}
\end{eqnarray}
where $V_{\phi}\equiv dV/d\phi$.
 Note that the above three equations are not independent. To solve these equations,
 we introduce the superpotential function $W(\phi)$,  which is
defined as:
\begin{eqnarray}
    A'=-\frac{\kappa^{2}}{n-2}W.
    \label{flat superpotential}
\end{eqnarray}
Substituting Eq.~(\ref{flat superpotential}) into Eqs.~(\ref{scalar flat Eom 1}) and
(\ref{scalar flat Eom 2}), we obtain Ref.~\cite{Liu:2012mia}:
\begin{eqnarray}
    \phi'&=&(1+c_{1}W^{2})W_{\phi},  \label{thick flat potential 1}  \\
     V&=&\frac{1}{2}(1+c_{1}W^{2})^{2}W^{2}_{\phi} - c_{2}W^{4} - c_{3}W^{2} - \frac{n-2}{2\kappa^{2}}\Lambda_{0},
\end{eqnarray}
where
\begin{eqnarray}
    && c_{1}=\frac{(n-4) }{2(n-2)^{2}} \zeta \kappa^{4},   \nonumber \\
    && c_{2}=\frac{ (n-1)(n-4) }{8(n-2)^{3}} \zeta \kappa^{6} ,  \\
    && c_{3}=\frac{(n-1) }{2(n-2)} \kappa^{2}.  \nonumber
    %\\
%    && W_{\phi}=\frac{dW}{d\phi},    \nonumber
\end{eqnarray}

\subsubsection{The case $\zeta=0$}
For the case $ \zeta = 0 $, we have  $c_1=c_2=0$.
In order to support a kink solution for the scalar filed, we first use the superpotential
$W = k v_0 \left(\phi -\frac{\phi^3}{3v_0^2} \right)$, for which the potential is
\begin{eqnarray}
   V(\phi)\! =\! -\frac{(n-1) k^2\kappa ^2}{18 (n-2) v_0^2}
             ( \phi^2 - v_0^2 )^2
             \!\left[ \!
                \phi^2 \! - \!4 v_0^2- \frac{9(n-2)}{(n-1) \kappa ^2} \!
             \right].\nonumber \\
\end{eqnarray}
The naked cosmological constant
$ \Lambda_0 = -\frac{4 (n-1)k^2 v_0^4 \kappa ^4}{9 (n-2)^2} $ is negative.
Substituting the superpotential into Eqs. (\ref{flat superpotential})
and (\ref{thick flat potential 1}), we obtain
\begin{eqnarray}
    \phi(y) &=& v_0 \tanh(ky),   \\
   \mathrm{e}^{2A(y)} &=& \mathrm{e}^{-\frac{v_0^2 \kappa^2}{3(n-2)}\tanh^2(k y)}
    [ \cosh(ky) ]^{ -\frac{4 v_0^2\kappa^2 } {3(n-2)} }.
\end{eqnarray}

Also we can take another superpotential $ W=$

$\!\!\!\!\!\!\!\!k\phi_0^2 \sin(\phi/\phi_0)$. The solution is
\begin{eqnarray}
   V(\phi) &=& \frac{1}{2}k^2\phi_0^2
               \left( 1 + \frac{n-1}{n-2}\kappa^2\phi_0^2\right)
              \cos^2(\phi/\phi_0),\\
   \phi(y) &=& 2 \mathrm{sign}(y) \phi_0 ~\arccos\left(\frac{1+\mathrm{e}^{-ky}}
                 {\sqrt{2+2\mathrm{e}^{-2ky}}}
            \right), \\
   \mathrm{e}^{2A(y)} &=& [\mathrm{sech}(ky)]^{\frac{2}{n-2}\kappa^2 \phi_0^2},\\
   \Lambda_0 &=& -\frac{n-1}{(n-2)^2}k^2\kappa^4 \phi_0^4 .
\end{eqnarray}
Note that the potential here is the Sine-Gordon potential and the scalar has a single kink-like configuration.
The naked cosmological constant $\Lambda_0$ is also negative.

%
%The potential is
%\begin{eqnarray}
%   V(\phi) = -\frac{ a^2 \left( (n-1)\kappa^2 \phi_{0}^{2}-n+2 \right)}
%            {2(n-2)\phi_{0}^{2}}
%            \sinh^2(\frac{\phi}{\phi_0}).
%\end{eqnarray}
%The corresponding cosmological constant is
%
%\begin{eqnarray}
%\Lambda_0=\frac{a^2 \kappa ^2}{(n-2) \phi_0^2}.
%\end{eqnarray}

\subsubsection{The case $\zeta \neq 0$}

In the last subsection we cannot get a physical brane solution for a usual $\phi^4$ potential for vanishing $\zeta$ with the superpotential method. However, for the case $ \zeta \neq 0 $, we can consider the usual $ \phi^{4} $ potential by
setting $ W=a \phi $. The potential turns to
\begin{eqnarray}
    V &=& b ( \phi^{2} - v_{0}^{2} )^2,
\end{eqnarray}
where
\begin{eqnarray}
    b &=& -\frac{ a^{4}(n-4)\zeta\kappa^{6}
                 \left[- 4a^{2}\zeta\kappa^{2}(n-4) + (n-2)(n-1) \right] }
               {8(n-2)^{4}},     \nonumber   \\
    v_{0}^{2} &=&  -\frac{2(n-2)^{2}}{a^{2}(n-4)\zeta\kappa^{4}} ,    \nonumber
\end{eqnarray}
and the corresponding cosmological constant is
\begin{eqnarray}
    \Lambda_{0} = \frac{n-1}{(n-4)\zeta}.
\end{eqnarray}
Here, we consider the case of $n>4$ and require $v_0^2>0$ and $b>0$, this leads to $\zeta<0$ and $\Lambda_{0}<0$.
The solutions of the scalar field and the warped factor are
\begin{eqnarray}
    \phi(y) &=& v_{0} ~ \mathrm{tanh}(ky),  \\
    \mathrm{e}^{2A(y)} &=& \left[ \cosh(ky) \right]^{ -\frac{ 2\kappa^{2} v_{0}^{2} } { n-2 } },
\end{eqnarray}
where
\begin{eqnarray}
    k=\sqrt{\frac{-(n-4)\zeta}{2}}\frac{\kappa^2 a^2}{n-2}.
\end{eqnarray}

If a trigonometric superpotential is used, the scalar field can be a single kink, double kink, or even
multi-kink, and there can be various kinds of structure of the brane. The warped factor $A(y)$ and the scalar
field are related with the extra dimension $y$ by
\begin{eqnarray}
    y\!&=&\!\int \!\! \frac{1}{\left(1+c_1 W^2\right) W_{\phi }} \, d\phi,  \label{yphi}    \\
    A(y)\!&=&\!-\frac{\kappa ^2}{n-2}\int \!\! \frac{W}{\left(1+c_1W^2\right) W_{\phi }} \, d\phi. \label{Aphi}
\end{eqnarray}

\begin{figure}
\begin{center}
\subfigure[$V(\phi,\zeta)$ with $q=3$]{\label{figure V zeta}
    \includegraphics[width=7cm]{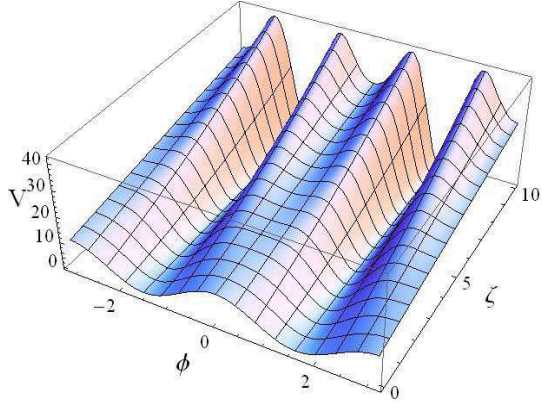}}
\subfigure[$V(\phi,q)$ with $\zeta=1$]{\label{figure V q}
    \includegraphics[width=7cm]{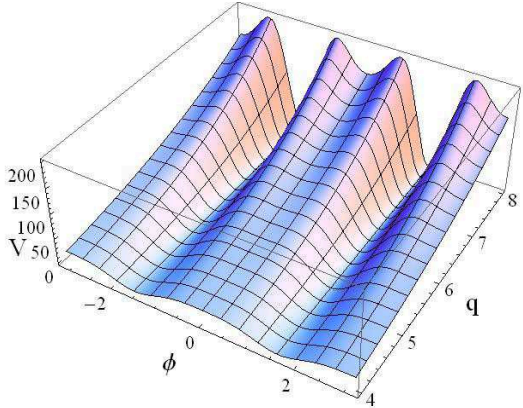}}
\end{center}
\caption{The shape of the scalar potential $V(\phi)$ for different values of $q$ and $\zeta$. The parameters are set to
         $n=5$, $\phi_0=1$, and $\kappa=1$. }
\end{figure}

\begin{figure}
\begin{center}
\subfigure[The shape of the scalar field]{\label{figure phiy zeta}
    \includegraphics[width=7cm]{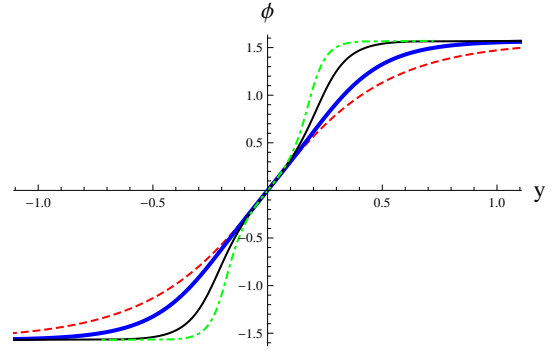}}
\subfigure[The shape of the energy density]{\label{figure rhoy zeta}
    \includegraphics[width=7cm]{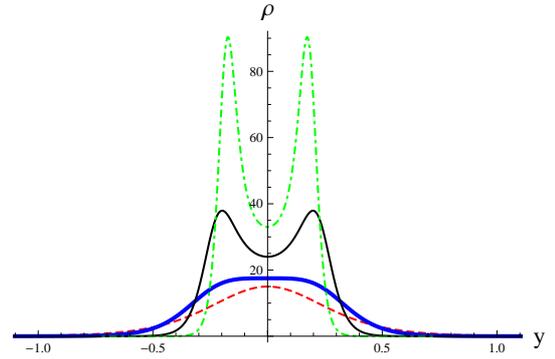}}
\end{center}
\caption{The shape of the scalar field and the energy density for different values of $\zeta$ for the flat thick brane. The parameters are set as $q=3$, $n=5$, $\phi_0=1$, $\kappa=1$,
         and $\zeta=0$ for the dashed red line, $ \zeta = \zeta_{c1}=\frac{5}{3}$ for the thick
         blue line, $\zeta=6$ for the thin black line, and $\zeta=12$ for the dotdashed green line.
          }
\end{figure}

\begin{figure}
\begin{center}
\subfigure[The shape of the scalar field]{\label{figure phiy q}
    \includegraphics[width=7cm]{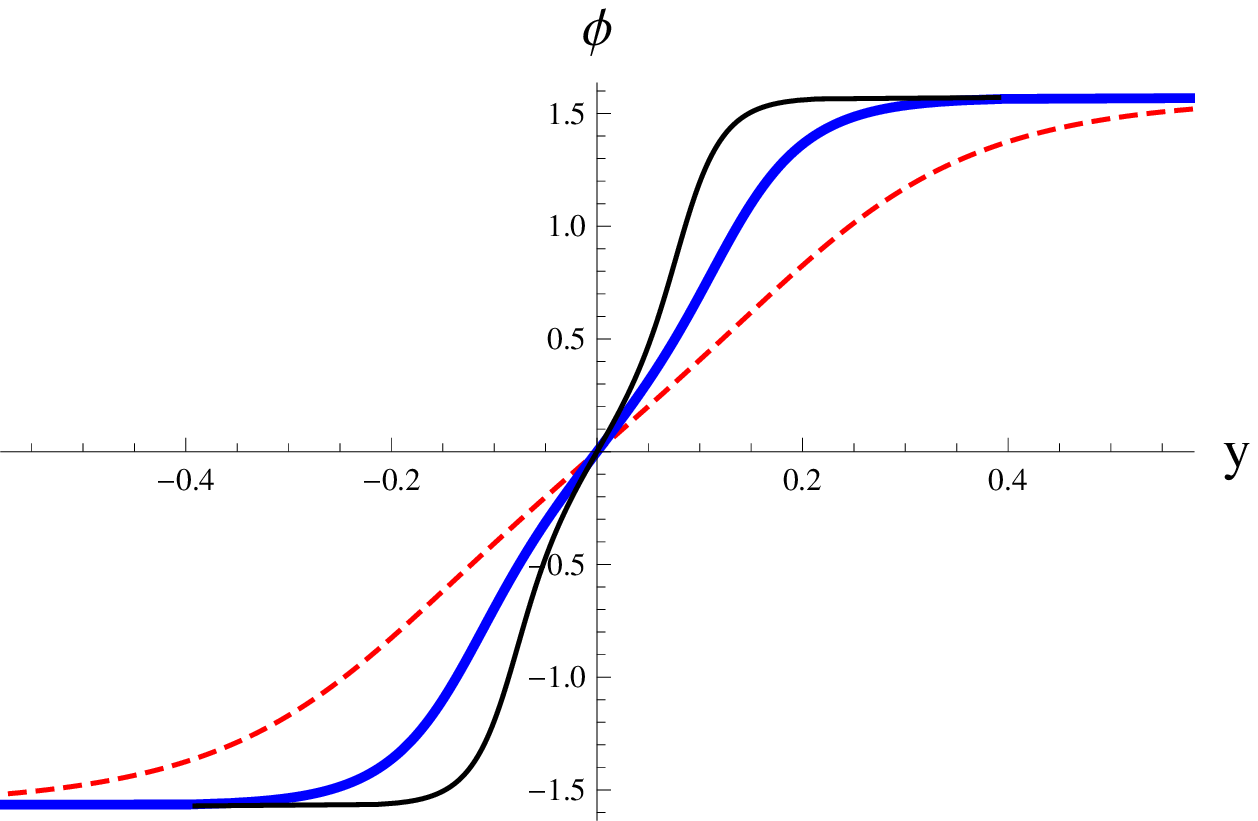}}
\subfigure[The shape of the energy density]{\label{figure rhoy k}
    \includegraphics[width=7cm]{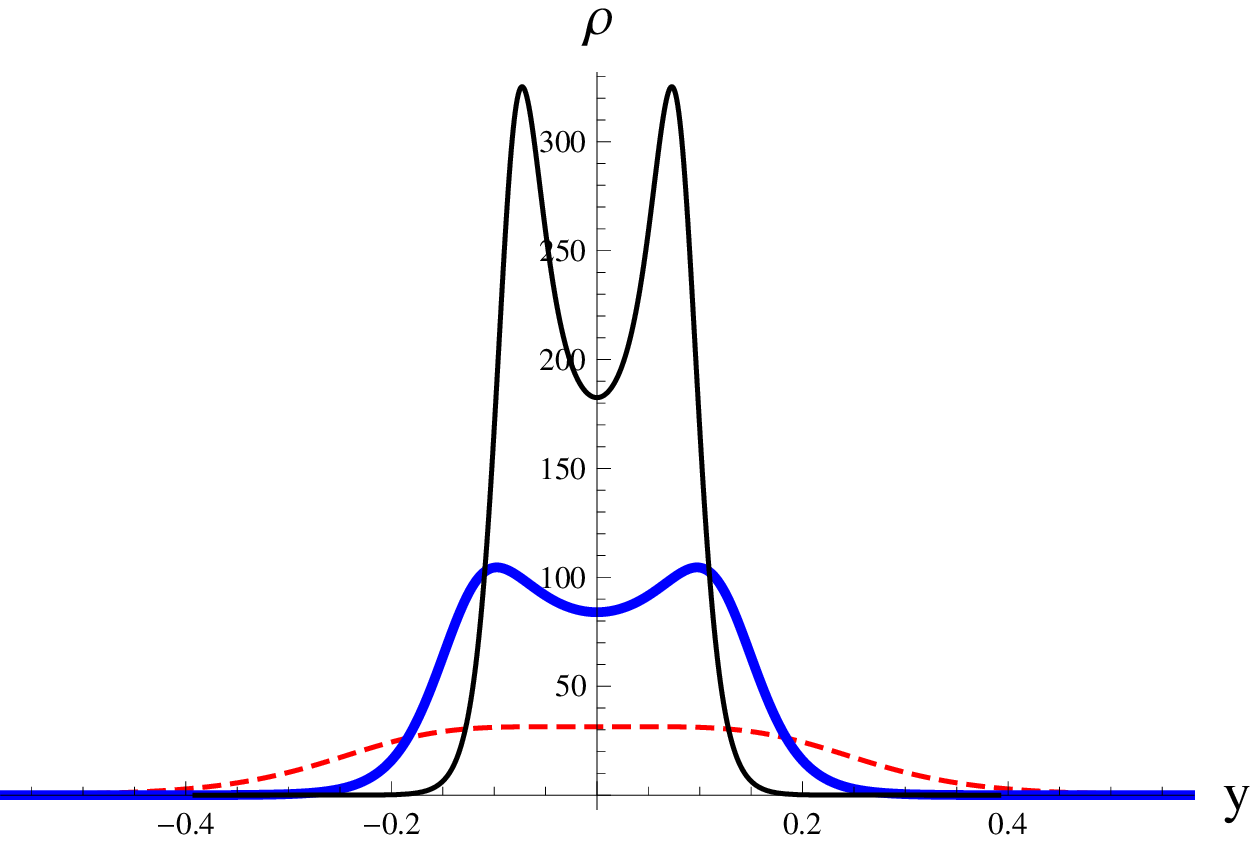}}
\end{center}
\caption{The shape of the scalar field and the energy density for different values of $q$ for the flat thick brane. The parameter $q$ is set to
         $q=4$ for the dashed red line, $q=6$ for the thick blue line, and $q=8$ for the thin black line.
         The other parameters are set as $\zeta=1$, $n=5$, $\phi_0=1$, and $\kappa=1$. }
\end{figure}

For the case $W(\phi)=q \phi_0 \sin \frac{\phi }{\phi_0} $, we
find that the parameters $\zeta$ and $q$ can
affect the structure of the brane. To see this, we first plot the scalar potential $V(\phi)$ in Figs. \ref{figure V zeta} and \ref{figure V q}, which show the influence of the parameters
$\zeta$ and $q$, respectively.
We can see that as  $\zeta$ and $q$ get larger, a fake vacuum of the scalar potential will
emerge, which is different from the case in general relativity, i.e. $\zeta=0$. So we can expect that the scalar field has a double kink solution, which
is shown in Figs. \ref{figure phiy zeta} and \ref{figure phiy q}; and the brane
is a double brane which can be seen from the energy density $\rho(y)=T_{MN}U^M U^N=-{T^{0}}_{0} =  \frac{1}{2}\phi'^2+V$ in Figs. \ref{figure rhoy zeta}
and \ref{figure rhoy k}.

The corresponding cosmological constant is
\begin{eqnarray}
\Lambda_0=-\frac{(n-1)q^2 \kappa ^4 \phi_0^4 \left[(n-4)q^2 \zeta\kappa^4 \phi_0^4+4(n-2)^2\right]}{4 (n-2)^4}.\nonumber \\
\end{eqnarray}

The condition that the single brane split into a double brane is
\begin{eqnarray}
    \rho''|_{y=0}>0,
\end{eqnarray}
i.e.,
\begin{eqnarray}
    \zeta >\zeta_{c1}\equiv\frac{ (n-2) \left[ 2(n-2)+(n-1)\kappa^2 \phi_0^2 \right] }
                {2(n-4)\kappa^4  \phi_0^2 q^2}.
\end{eqnarray}
From Fig. \ref{figure rhoy k}, it can be seen that with the increase of the parameter $\zeta$, the brane become fatter. When $\zeta$ reaches the critical value $\zeta_{c1}$, there will be a wide platform around the brane location. When $\zeta>\zeta_{c1}$, there will be a minimum for the energy density at the center of the brane and two sub-branes appear. Such brane with inner structure may support resonant KK modes for various bulk matter fields.

\subsection{AdS thick Brane}

Now we consider the AdS thick brane, for which the line-element is also assumed as (\ref{AdS metric}) and the EoMs read as
\begin{eqnarray}
    \label{AdS EoM1}
     &&\left[(n-4)\zeta H^{2}\mathrm{e}^{-2A} + (n-4)\zeta A'^{2} +2 \right]
    ( A''-H^{2}\mathrm{e}^{-2A}) \nonumber\\
   && = -\frac{2\kappa^{2}}{n-2}\phi'^{2}, \\
    && (n-1) A'^{2}\left[ (n-4)\zeta A'^{2}+2(n-4)H^{2}\zeta\mathrm{e}^{-2A}+ 4 \right]  \nonumber\\
     &&   +4(n-1)H^{2}\mathrm{e}^{-2A}
     + (n-4)(n-1)\zeta H^{4}\mathrm{e}^{-4A} + 4\Lambda_{0} \nonumber\\
    && =\frac{8\kappa^{2}}{n-2}\Big(\frac{1}{2}\phi'^{2}-V\Big),    \\
    && \phi'' + (n-1)A'\phi'-\frac{\partial V}{\partial\phi} = 0.
\end{eqnarray}

\subsubsection{The case $\zeta=0$}
For the case $\zeta=0$, we consider the Sine-Gordon potential
\begin{eqnarray}
    V(\phi) = -\frac{1}{16} (n-2) \phi_0^2 \, \left[ \mathrm{cos} \left( \frac{4k \phi }{\phi_0} \right)+1 \right] .
\end{eqnarray}
Then we get the following solution
\begin{eqnarray}
    \mathrm{e}^{2A(y)} &=& \mathrm{cosh}^2(k y),    \\
    \phi (y) &=& \frac{ \phi_0 }{k} ~ \mathrm{arctan}\left(\mathrm{tanh}\left(\frac{k y}{2}\right)\right),
\end{eqnarray}
where
\begin{eqnarray}
    \phi_0 &=& \frac{2\sqrt{(n-2)(H^2-k^2)}}{ \kappa }.
\end{eqnarray}
The cosmology constant is
\begin{eqnarray}
    \Lambda_0 &=& -(n-1)k^2,
\end{eqnarray}
which is negative. Therefore, the AdS thick brane is embedded in an asymptotic AdS spacetime.
It is interesting to note that the single kink scalar connects the adjacent locations of the
 extrema of the scalar potential, and the energy density
\begin{eqnarray}
    \rho = -\frac{1}{8}(n-2)\phi_0^2 \mathrm{sech}^2(ky)
\end{eqnarray}
is negative. This is very different from the case of flat branes.

\subsubsection{The case $\zeta \neq 0$}
\begin{figure}
\begin{center}
\subfigure[The shape of the potential]{\label{figure V AdS2}
    \includegraphics[width=7cm]{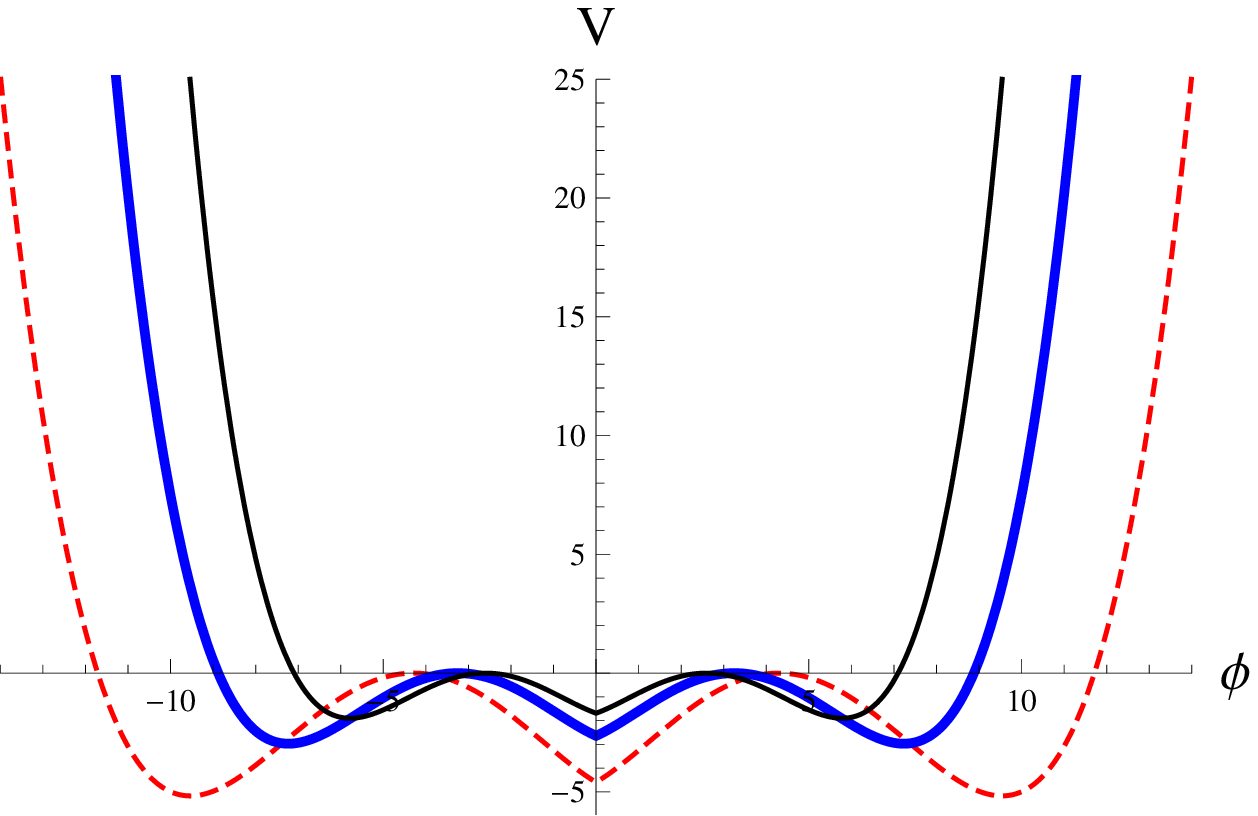}}
\subfigure[The shape of the scalar field]{\label{figure phi AdS2}
    \includegraphics[width=7cm]{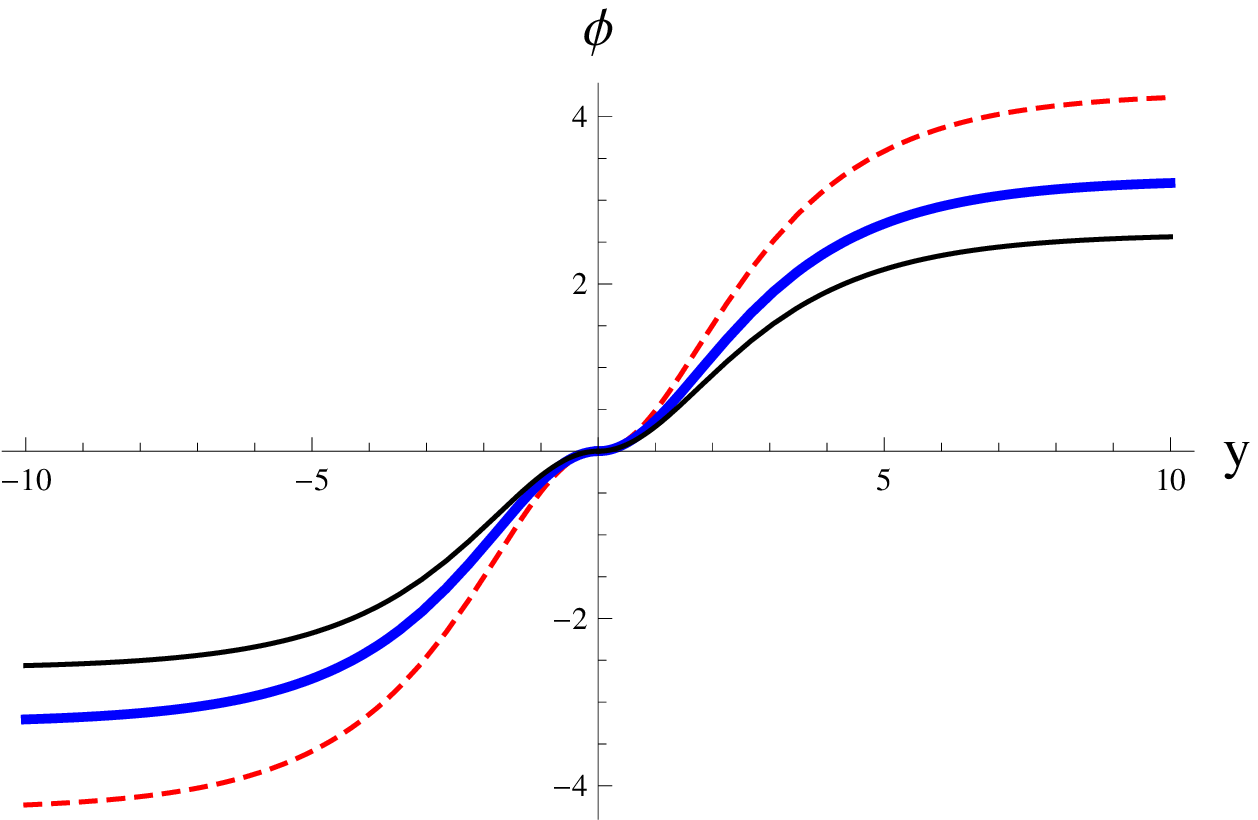}}
\subfigure[The shape of the energy density]{\label{figure rho AdS2}
    \includegraphics[width=7cm]{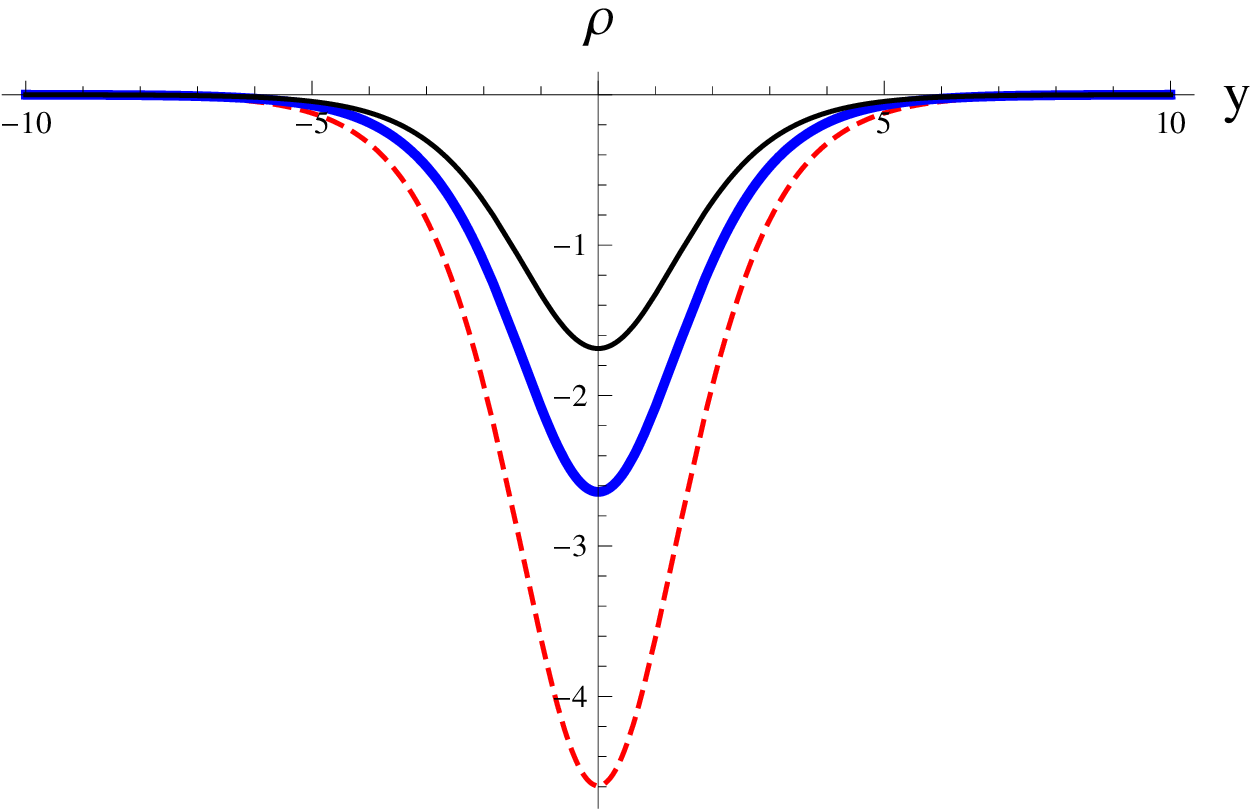}}
\end{center}
\caption{The shape of the potential, the scalar field, and the energy density
         for different values of $k$ for the AdS thick brane. The parameter $\zeta$ is set to
         $\zeta=-1.0$ for the dashed red line, $\zeta=-1.5$ for the thick blue line,
         and $\zeta=-2.0$ for the thin black line.
         The other parameters are set as  $n=5$, $H=1.0$, $k=0.5$, and $\kappa=1.0$. }
\end{figure}

For the case $\zeta<0$, we find a solution for a $\phi^4$ model:
\begin{eqnarray}
    \mathrm{e}^{A(y)} &=& \sqrt{ -\frac{1}{2} H^{2} \zeta (n-4) } \cosh(ky),    \\
    \phi (y) &=& \mathrm{sign}(y) \phi_{0}(1-\mathrm{sech}(ky)),    \\
    V(\phi) &=& b [ (| \phi | - \phi_{0})^{2} - v_{0}^{2} ]^2 - b v_0^4,
\end{eqnarray}
in which
\begin{eqnarray}
    %b &=& -\frac{k^4 (n-4) \left(6-5 n+n^2\right) \zeta  \left(2+k^2 (n-4) \zeta \right)^2}{2 (n-2)^2 \kappa ^2 \left[\frac{2+k^2 (n-4) \zeta }{\kappa }\right]^4},  \\
    b &=& -\frac{ (n-3)(n-4)    \zeta \kappa^2 k^4}
                {2 (n-2)  \left[{2+k^2 (n-4) \zeta }\right]^2},  \\
    \phi_{0} &=& \frac{1}{k{\kappa}}\left| {2+ (n-4)\zeta k^2} \right| \sqrt{-\frac{(n-2) }{2(n-4) \zeta }},  \\
    v_{0}^2 &=& \frac{n-2}{n-3} \phi_0^2 .
\end{eqnarray}
The energy density $\rho(y)$ of the system is
\begin{eqnarray}
    \rho(y)
     &=&
   \frac{\phi_0^2}{4k^2} [ (k^4-n+2)\mathrm{cosh}(2ky)  -k^2-1 ]\mathrm{sech}[k y]^4 . \nonumber \\
\end{eqnarray}
From the above expression of the energy density we can see that
the brane tension of the AdS brane is negative, which is different from the cases of flat and dS branes. The solution of the scalar field is a double kink. At the boundaries of the extra dimension, the scalar field $ \phi \rightarrow \pm\phi_0$, which are locations of the extrema of the scalar potential, but not the locations of the minima. This is the reason why the energy density is negative. The shape of the scalar field, the potential
and the energy density are shown in Figs. \ref{figure V AdS2}-\ref{figure rho AdS2}.

The cosmology constant is
\begin{eqnarray}
    \Lambda_0 =-\frac{1}{4} k^2 (n-1) \left[(n-4) \zeta k^2+4 \right].
\end{eqnarray}

\begin{figure}
\begin{center}
\subfigure[The shape of the scalar field]{\label{figure AdS numeric phi}
    \includegraphics[width=7cm]{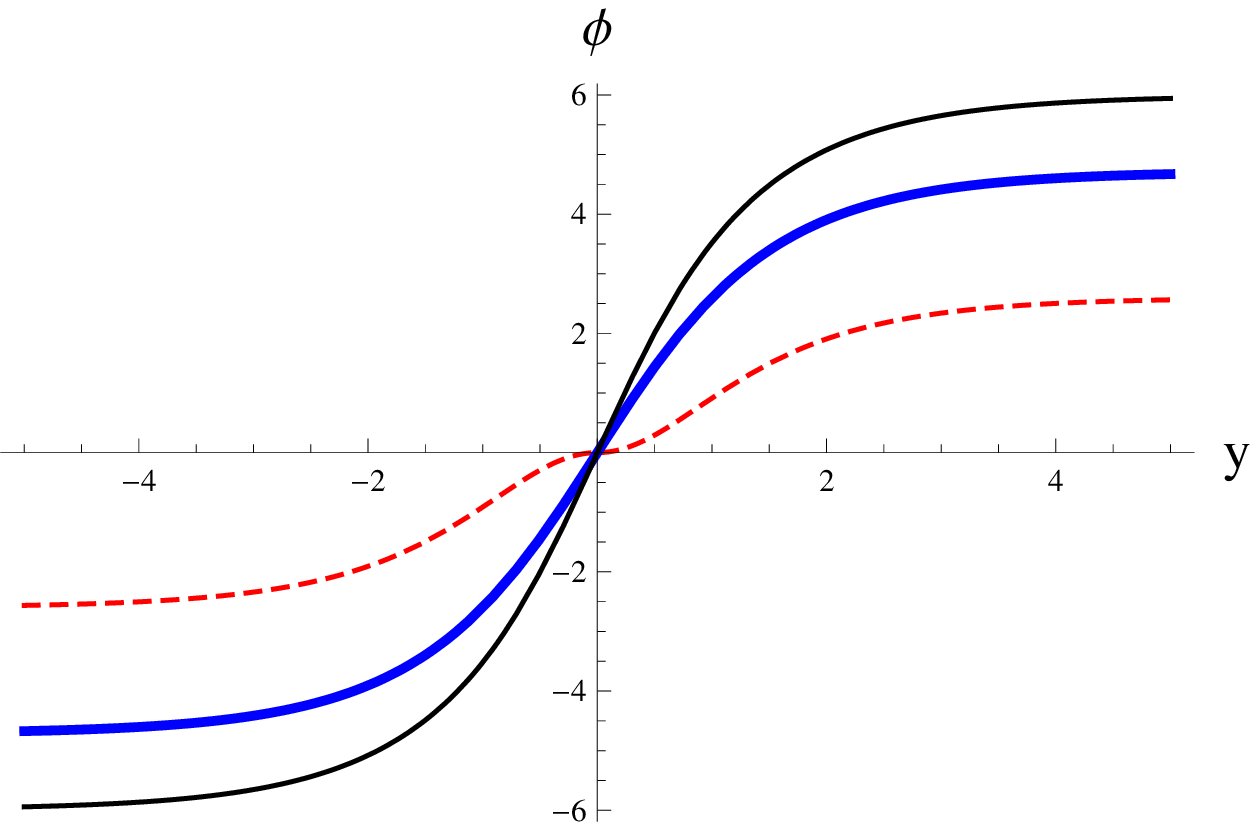}}
\subfigure[The shape of the energy density]{\label{figure AdS numeric rho}
    \includegraphics[width=7cm]{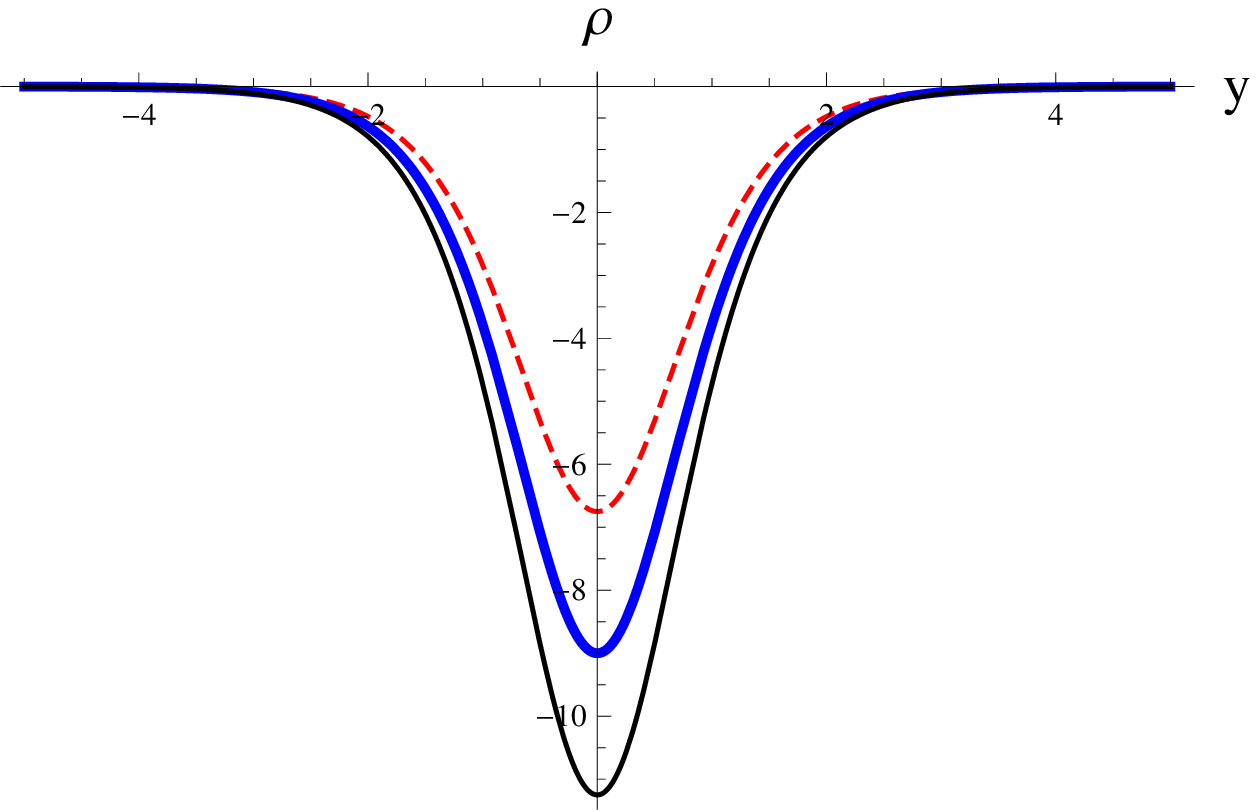}}
\end{center}
\caption{The shape of the scalar field and energy density for different values of $\zeta$ for the  AdS thick brane.
         The parameter $\zeta$ is set to $\zeta = -\frac{2}{(n-4)H^2}$ for the dashed red line, $\zeta=0$ for the thick
         blue line, and $\zeta = \frac{2}{(n-4)H^2}$ for the thin black line.
         The other parameters are set to $n=5$, $\kappa=1.0$, $k=1.0$, and $H=2.0$. }\label{figure dS zeta}
\end{figure}

For the following warped factor:
\begin{eqnarray}
    \mathrm{e}^{2A(y)}
    &=& \mathrm{cosh}^{-2}(ky),
\end{eqnarray}
the numerical solutions of the scalar field and energy density are shown in Figs. \ref{figure AdS numeric phi}
and \ref{figure AdS numeric rho} for different values of $\zeta$. From Eq.(\ref{AdS EoM1}),
$\phi'^2(y)\geq0$ implies
\begin{eqnarray}
    \zeta \geq -\frac{2}{(n-4)H^2},
\end{eqnarray}
which yields
\begin{eqnarray}
    \rho\leq0,  ~~  \rho''|_{y=0}>0.
\end{eqnarray}
Therefore, the brane is a single brane with negative tension. It is interesting to note
that, the scalar is a double kink when $\zeta = -\frac{2}{(n-4)H^2}$. This double-kind structure could contribute to the resonant structure of bulk fermions.
The cosmology constant is
\begin{eqnarray}
    \Lambda_0 =-\frac{1}{4} (n-1) k^2 \left[(n-4) \zeta k^2+4 \right].
\end{eqnarray}

\subsection{dS thick brane}
In order to simplify the EoMs in the case of dS brane, we introduce the conformal coordinates.
The line-element is assumed as
\begin{equation}
    ds^2 = e^{2A} [-dt^{2}+\mathrm{e}^{-2Ht}\delta_{ij} dx^{i} dx^{j}+ dz^{2} ].
    \label{conformal dS metric}
\end{equation}
Then the EoMs are turned out to be
 \begin{eqnarray}
  &&   \left[(n-4)\zeta\mathrm{e}^{-2A} (A'^{2}-H^{2}) +2 \right](A''+H^{2}-A'^2) \nonumber\\
        && = -\frac{2\kappa^{2}}{n-2}\phi'^{2}, \\
   &&  (n-1)(A'^{2}-H^{2})\left[ (n-4)\zeta\mathrm{e}^{-2A}(A'^{2}-H^{2})+4 \right]\nonumber\\
   &&  + 4\mathrm{e}^{-2A}\Lambda_0
     =\frac{4\kappa^{2}}{n-2}(\phi'^{2}-2\mathrm{e}^{2A}V), ~~~~~~~~   \\
   &&  \mathrm{e}^{-2A}\left[ \phi''+(n-2)A'\phi'\right]-\frac{\partial V}{\partial\phi} = 0,
\end{eqnarray}
where the prime denotes the derivative with respect to $z$.

\subsubsection{The case $\zeta=0$}
For the case $\zeta=0$, the EoMs reduce to the ones in general relativity. The solution for $n=5$ has
been given in Ref.~\cite{Liu:2009ve,Goetz:1990,Gass:1999gk} in general relativity. For arbitrary $n$, we consider the following potential
\begin{eqnarray}
    V(\phi) &=& \frac{(n-2)}{2p \kappa^2 }
    \left[ (n-2)p+1 \right] H^{2} \cos^{2(1-p)}\left(\frac{2\phi}{\phi_0}\right),
\end{eqnarray}
where the parameter $p$ satisfies $0<p<1$.
The solution of the warped factor and scalar field is
\begin{eqnarray}
    \mathrm{e}^{2A(z)}
    &=& \mathrm{cosh}^{-2p}\left(\frac{Hz}{p}\right),  \\
    \phi (z) &=&
    \phi_{0} \mathrm{arctan}\left(\mathrm{tanh}\left(\frac{Hz}{2p}\right)\right),
\end{eqnarray}
with
\begin{eqnarray}
    \phi_0 = \frac{2}{ \kappa }\sqrt{(n-2)p(1-p)}.
\end{eqnarray}
The corresponding naked cosmology constant is
\begin{eqnarray}
    \Lambda_0 = 0.
\end{eqnarray}
Since $R_{MN}(z\rightarrow\infty)\rightarrow 0$, the effective cosmological constant
is also zero. So the dS thick brane is embedded in an $n$-dimensional Minkowski spacetime.
The energy density is given by
\begin{eqnarray}
    \rho = \frac{3(1+p)H^2}{p\kappa^2} \mathrm{sech}^{2(1-p)}\left(\frac{H}{p}z\right).
\end{eqnarray}

\subsubsection{The case $\zeta \neq 0$}
\begin{figure}
\begin{center}
\subfigure[The shape of the scalar field]{\label{figure dS H phi}
    \includegraphics[width=7cm]{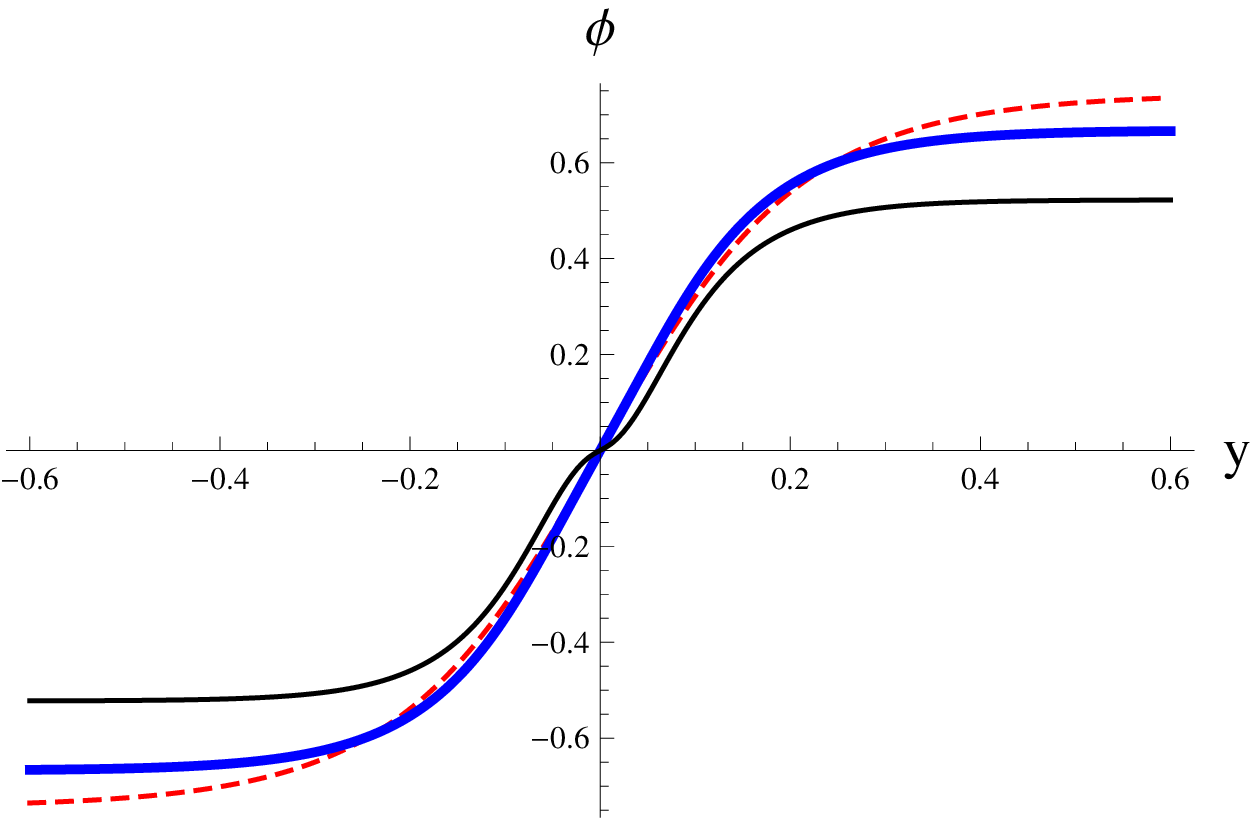}}
\subfigure[The shape of the energy density]{\label{figure dS H rho}
    \includegraphics[width=7cm]{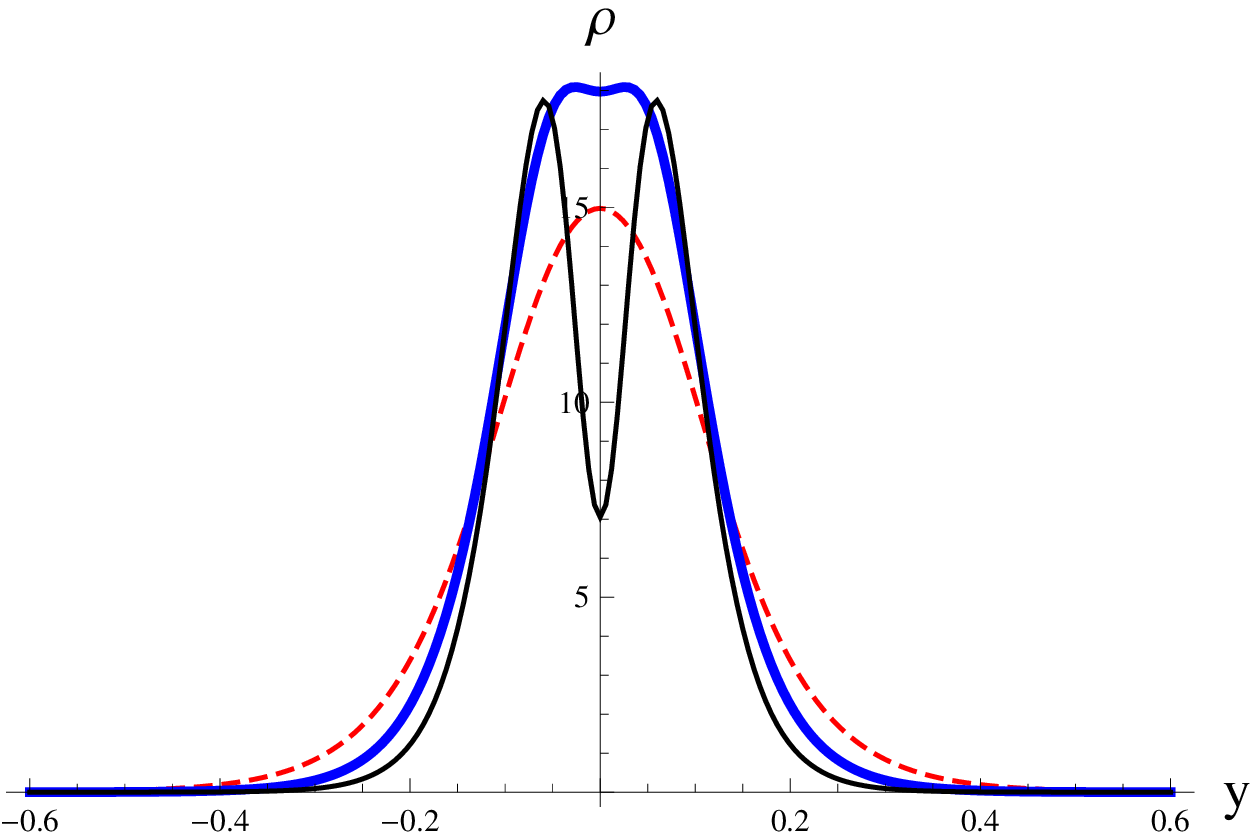}}
\end{center}
\caption{The shape of the scalar field and energy density for different values of $H$  for the  dS thick brane.
         The parameter $H$ is set to $H=0.8$ for the dashed red line, $H=1.1$ for the thick
         blue line, and $H=1.4$ for the thin black line.
         The other parameters are set to $n=5$, $p=0.1$, $\Lambda_0=0.0$, $\kappa=1.0$ and $\zeta=1.0$. }\label{figure dS H}
\end{figure}

For the case $\zeta \neq 0$, it is hard to find a closed solution. For the following warped factor:
\begin{eqnarray}
    \mathrm{e}^{2A(z)}
    &=& \mathrm{cosh}^{-2p}\left(\frac{Hz}{p}\right),
\end{eqnarray}
the numerical solutions of the scalar field and energy density are shown in Figs. \ref{figure dS H}
and \ref{figure dS zeta} for different values of $H$ and $\zeta$, respectively.

\begin{figure}
\begin{center}
\subfigure[The shape of the scalar field]{\label{figure dS zeta phi}
    \includegraphics[width=7cm]{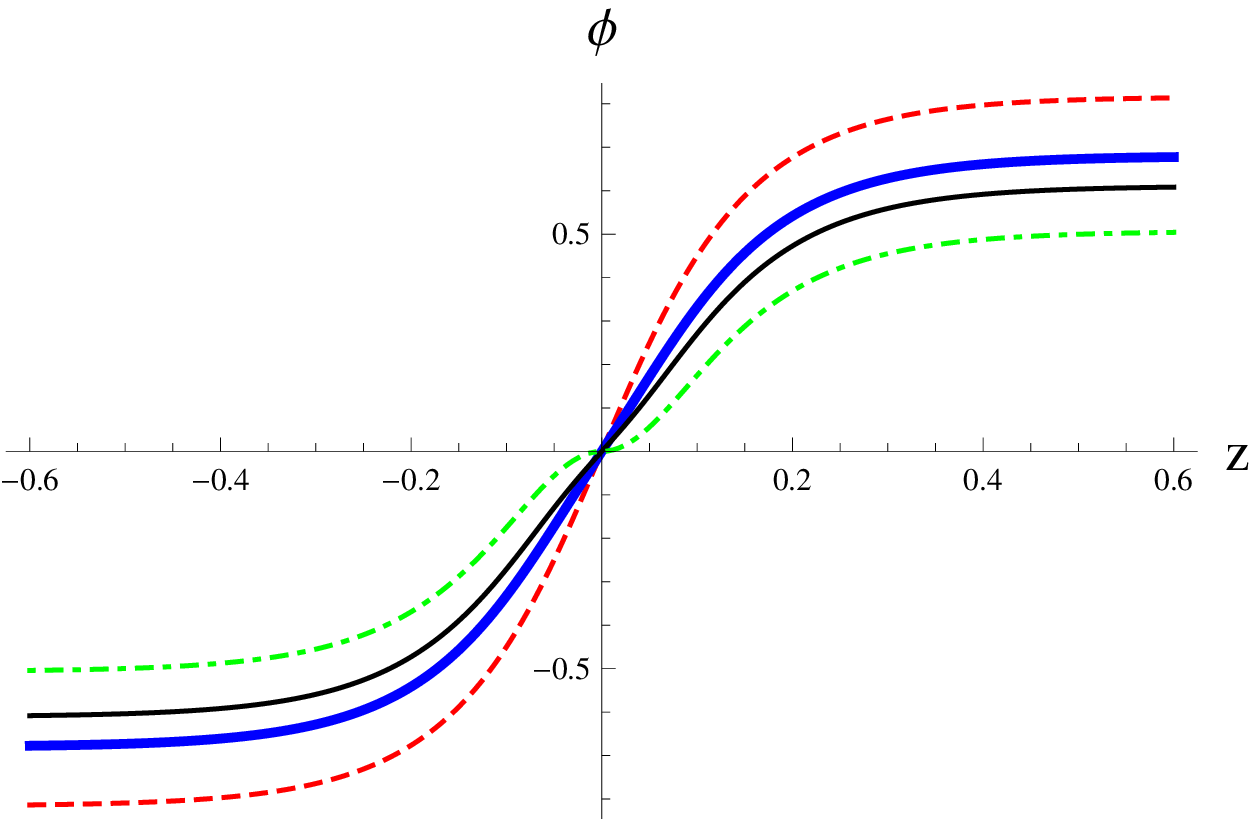}}
\subfigure[The shape of the energy density]{\label{figure dS zeta rho}
    \includegraphics[width=7cm]{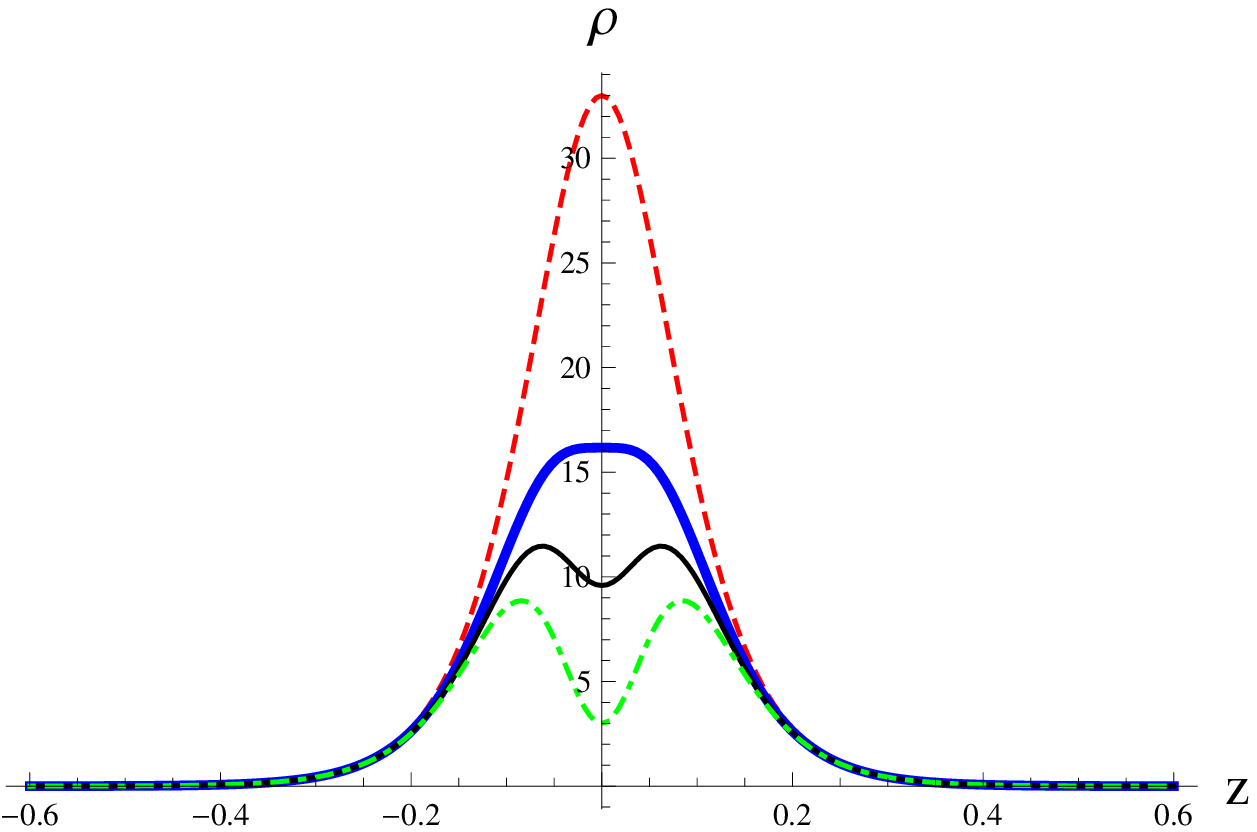}}
\end{center}
\caption{The shape of the scalar field and energy density for different values of $\zeta$ for the  dS thick brane. The parameters are set to $n=5$, $\Lambda_0=0.0$, $\kappa=1.0$, $p=0.1$, $H=1.0$, and $\zeta=0$ for the dashed red line, $\zeta=\zeta_{c2}=1.12195$ for the thick
         blue line, $\zeta=1.56$ for the thin black line, and $\zeta=\frac{2}{(n-4)H^2}=2$ for the dotdashed green line.
          }\label{figure dS zeta}
\end{figure}

We can see that as  $H$ and $\zeta$ get larger, the scalar field turns to a double kink (see
Figs. \ref{figure dS H phi} and \ref{figure dS zeta phi}); and the brane splits into two sub-branes, which can be seen from the energy density $\rho$ in Figs. \ref{figure dS H rho} and \ref{figure dS zeta rho}. This is different from the case of $\zeta=0$.

The condition that the single brane splits into the double brane is
\begin{eqnarray}
    \zeta > \zeta_{c2} \equiv\frac{ 2(n-2)p +4 }
                {(n-4)\left[ (n-4)p+4 \right]H^2}.
\end{eqnarray}

\section{Conclusion}

In this paper, we generalized the Minkowski brane models in five-dimensional
critical gravity in Ref. \cite{Liu:2012mia} to warped ones in $n$ dimensions. For thin brane models in arbitrary dimensional critical gravity theory, the Gibbons-Hawking surface term and  the junction conditions were derived. It was found that for the special case of flat, AdS, and dS thin branes the $C^2$ term in the action has no contribution to these junction conditions.
The solutions for both thin and thick branes were obtained at
the critical point $\alpha=-\frac{n}{4(n-1)}\beta$.

We found that the combination of the parameters $\beta$ and $\gamma$ in the action (\ref{LG1}), i.e., $\zeta\equiv\beta(n-2)-4\gamma(n-3)$, has nontrivial effect on the brane solutions.
All the flat, AdS, and dS thin branes are embedded in an AdS$_{n}$
spacetime, and the effective cosmological constant $\Lambda$ of the AdS$_{n}$ spacetime equals the naked one $\Lambda_0$ only when the combine coefficient  $\zeta=0$. The naked cosmological constant and brane tension can be positive, zero, and negative, depending
on the value of $\zeta$. Following the procedure in Ref. \cite{Randall:1999ee}, we reduce the $n$-dimensional critical gravity to the $(n-1)$-dimensional critical gravity on the brane, and the mass hierarchy problem was also solved in the higher-order braneworld model in the critical gravity.

For the thick flat branes, when $\zeta=0$, we got two analytical solutions, both of which describe a single brane generated by a kink-like scalar. When $\zeta\neq0$, the analytical and numerical solutions were obtained. It was found that the brane will split into a double brane when one of the parameters $\zeta$ and $k$ is larger than its critical value.
Such brane with inner structure may support resonant KK modes for various bulk matter fields.
All these flat branes are embedded in an $n$-dimensional AdS spacetime.

For the thick AdS branes, the scalar connects the adjacent locations of the
extrema of the scalar potential, and the energy density is negative. This is very different from the cases of flat and dS branes. The scalar can have single or double kink configuration, but the brane has no inner structure. These AdS branes are also embedded in an $n$-dimensional AdS spacetime.

For the thick dS branes, when $\zeta=0$ and the scalar potential is taken as the Sine-Gordon one, the brane has positive energy density but has no inner structure. When $\zeta\neq0$, the inner structure of the dS brane will appear when the parameter $\zeta$ or $H$ is larger than its critical value. The energy density of the brane system is positive for any $\zeta$. These dS branes are embedded in an $n$-dimensional Minkowski spacetime.

{
We have investigated the branes with maximally symmetry, where
dS brane can be considered as a $(n-1)$-dimensional exponentially expanding
universe. We can also consider a FRW brane and reconsider the cosmological
constant problem and inflation in the frame of brane cosmology \cite{Wands:2006tz}. Moreover, we can also consider higher co-dimension branes, but the equations of motion will be forth-order differential ones even if critical condition (\ref{CriticalCondition}) is satisfied. It will be difficult to solve them analytically.
}

\begin{acknowledgements}
This work was supported by the National Natural Science Foundation of China (Grants No. 11075065 and No. 11375075), and
the Fundamental Research Funds for the Central Universities (Grants No. lzujbky-2013-18 and
No. lzujbky-2013-227).
\end{acknowledgements}

%\begin{acknowledgements}
%If you'd like to thank anyone, place your comments here
%and remove the percent signs.
%\end{acknowledgements}

% BibTeX users please use one of
%\bibliographystyle{spbasic}      % basic style, author-year citations
%\bibliographystyle{spmpsci}      % mathematics and physical sciences
%\bibliographystyle{spphys}       % APS-like style for physics
%\bibliography{}   % name your BibTeX data base

%\bibliographystyle{spphys}
%\bibliography{E:/jabrefdatabase/jabrefdatabase}

\end{document}